\documentclass[12pt]{book}
\usepackage{sussp} 
\usepackage{graphicx}     
\usepackage{makeidx}
\makeindex

\begin{document}

\addtolength{\voffset}{2cm}


\thispagestyle{empty}
\begin{flushright}
CERN-TH/2002-018\\
February 2002 
\end{flushright}

\vspace*{1.5cm}
\centerline{\Huge\bf Heavy Quark Theory}
\vspace*{2cm}
\centerline{{\sc Gerhard Buchalla}}
\bigskip
\centerline{\sl Theory Division, CERN, CH-1211 Geneva 23,
                Switzerland}
 
\vspace*{1.5cm}
\centerline{\bf Abstract}
\vspace*{0.3cm}
\noindent 
These lectures describe the most important theoretical
methods in $b$-physics. We discuss the formalism of
effective weak Hamiltonians, heavy quark effective theory,
the heavy quark expansion for inclusive decays of
$b$-hadrons and, finally, the more recent ideas of
QCD factorization for exclusive nonleptonic $B$ decays.
While the main emphasis is put on introducing the
basic theoretical concepts, some key applications in
phenomenology are also presented for illustration.

\vspace*{3cm}
\centerline{\it Lectures given at the}
\centerline{\bf 55th Scottish Universities Summer School in Physics}
\centerline{\bf Heavy Flavour Physics}
\centerline{\it University of St. Andrews, Scotland}
\centerline{\it 7 -- 23 August 2001}

\vfill

 
\newpage
\pagenumbering{arabic}


\headings{Heavy Quark Theory} 
{Heavy Quark Theory}
{Gerhard Buchalla}
{Theory Division, CERN, CH-1211 Geneva 23, Switzerland}

\section{Preface}

The dedicated study of $b$-flavoured hadrons has developed
into one of the most active and most promising areas of experimental
high-energy physics. The detailed investigation of $b$ decays at the
$B$-meson factories and at hadron colliders will probe the flavour
physics of quarks with unprecedented precision. To fully exploit
this rich source of data a systematic theoretical approach is necessary.
The required field theoretical tools are the subject of these
lectures.

We shall discuss the construction of effective weak Hamiltonians,
introducing the operator product expansion (OPE) and the renormalization
group (RG) and presenting the effective Hamiltonians for nonleptonic
$\Delta B=1$ and $\Delta B=2$ transitions as examples.
The subsequent chapter is devoted to heavy-quark effective theory (HQET).
It explains the basic formalism as well as the application to
heavy-light and heavy-heavy currents, discussing the $B$-meson decay
constant $f_B$ and the exclusive semileptonic decay $B\to D^*l\nu$,
respectively.
Inclusive $b$ decays and the heavy-quark expansion (HQE) are treated next,
in particular the general formalism, the issue of quark-hadron duality,
the theory of $b$-hadron lifetimes and of inclusive semileptonic decays.
Finally, we discuss QCD factorization for exclusive hadronic $B$ decays,
focussing on $B\to D\pi$ and $B\to\pi\pi$. We conclude with a short
summary.

The effective-Hamiltonian framework is the oldest and most general of
the methods we shall discuss. It dates back, more or less, to the
beginnings of the standard model itself. HQET and HQE are later
developments that have been established in the second half of the eighties
and at the beginning of the nineties. The last topic, QCD factorization
for exclusive $B$ decays is the most recent. It is the least well established
among these methods and it continues to be studied and developed in
further detail.

We would like to mention a very short selection of literature, which
we hope will be helpful to obtain further information on the
various subjects related to the contents of these lectures.
Very useful resources are the BaBar Physics Book 
(Harrison \& Quinn 1998) and the Fermilab $B$ Physics Report
(Anikeev et al. 2002). They collect nice reviews on both theoretical
and experimental topics in $B$ physics.
A textbook more specifically directed towards theoretical
heavy quark physics is the work of Manohar \& Wise (2000).
Review articles on particular subjects are 
(Buchalla, Buras \& Lautenbacher 1996) on effective weak Hamiltonians,
(Neubert 1994) on HQET and (Bigi et al. 1994; 
Bigi, Shifman and Uraltsev 1997) on HQE.

\section{Introduction and overview}

\subsection{Motivation}

In the following chapters we will study the theoretical tools
to compute weak decay properties of heavy hadrons. To put the
formalism into perspective, we start by recalling the main motivation
for this subject.

\begin{figure}
[htbp]
\centering
\includegraphics[width=15cm,height=3cm]{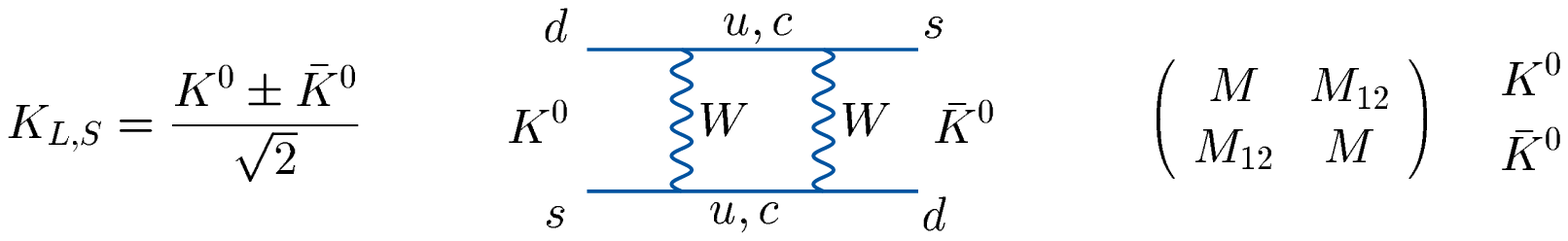}
\caption{$K^0$ -- $\bar K^0$ mixing.}
\label{fig:kkbar}
\end{figure}

The central goal is the investigation of flavour physics, the most
complicated sector in our understanding of fundamental interactions.
A good example is give by particle-antiparticle mixing, as first 
studied with neutral kaons. The strong interaction eigenstates
$K^0$ ($\bar sd$) and $\bar K^0$ ($\bar ds$) can be transformed into
each other through second order weak interactions, which leads to a
tiny off-diagonal entry $M_{12}$ in the mass matrix
(Fig. \ref{fig:kkbar}).
The physical eigenstates are $K_{L,S}=(K^0\pm \bar K^0)/\sqrt{2}$.
Their mass difference $\Delta m_K=m_L-m_S$ is given by
$2|M_{12}|$ and reads
\begin{equation}
\label{delmk}
\frac{\Delta m_K}{m_K}\approx\frac{G^2_F f^2_K B_K}{6\pi^2}
\left|V_{cs} V_{cd}\right|^2 m^2_c = 7\cdot 10^{-15}
\end{equation}
where the number on the right-hand side is the experimental value.
The theoretical expression is derived neglecting the third
generation of quarks and CP violation, which is a valid approximation
for $\Delta m_K$. The factors $f^2_K B_K$ ($B_K\approx 1$ for the
purpose of a first estimate) account for the binding of the quarks
into mesons. A crucial feature of (\ref{delmk}) is the 
Glashow-Iliopoulos-Maiani (GIM) cancellation mechanism:
The orthogonality of the quark mixing matrix $V_{ij}$ ($i=u,c$;
$j=d,s$) leads to a cancellation among the various contributions
with intermediate up and charm quarks 
(symbolically the amplitude has the form $(uu)-(uc)-(cu)+(cc)$).
For $m_u=m_c$ this cancellation would be complete, giving $\Delta m_K=0$.
Still, even for $m_u\not= m_c$, the contributions from virtual
momenta $k\gg m_c$, $m_u$ cancel since both $m_u$ and $m_c$
are negligible in this case. What is left is a characteristic effect
proportional to $m^2_c$, the up-quark contribution being subleading
for $m_u\ll\Lambda_{QCD}\ll m_c$. This circumstance allowed
Gaillard and Lee in 1974 to correctly estimate the charm-quark
mass $m_c\approx 1.5$ GeV, before charm was eventually discovered in the
Fall of the same year.

In a similar way the discovery of $B_d$ -- $\bar B_d$ mixing by the ARGUS
collaboration (Albrecht et al. 1987) proved to be another milestone in 
flavour physics. In full analogy to $K$ -- $\bar K$ mixing we have
\begin{equation}
\label{delmb}
\frac{\Delta m_B}{m_B}\approx\frac{G^2_F f^2_B B_B}{6\pi^2}
\left|V_{tb} V_{td}\right|^2 M^2_W S(\frac{m^2_t}{M^2_W}) = 6\cdot 10^{-14}
\end{equation}
where now the top-quark contribution is completely dominant.
The unexpectedly large value observed by ARGUS provided the first evidence
that the top-quark mass ($m_{t,pole}=176$ GeV) was comparable to the weak
scale and in any case much heavier than anticipated at the time.

These examples show very nicely the ``flavour'' of flavour physics:
Precision observables are sensitive probes of high-energy scales
and yield crucial insights into the fundamental
structure of weak interactions.
At the same time we see that hadronic effects manifest in quantities such as
$f_B$, $B_B$, and strong interactions of the participating quarks in general,
play an important role. Their understanding is necessary to reveal the
underlying flavour dynamics and is the main subject of heavy quark theory.

\begin{figure}
[htbp]
\centering
\includegraphics[width=9cm,height=6cm]{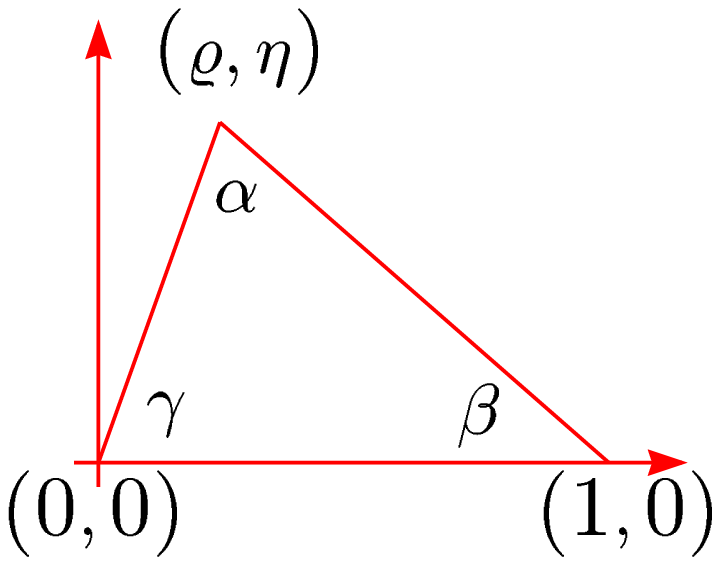}
\caption{Definition of the unitarity triangle.}
\label{fig:utdef}
\end{figure}

\begin{figure}
[htbp]
\centering
\includegraphics[width=15cm,height=10cm]{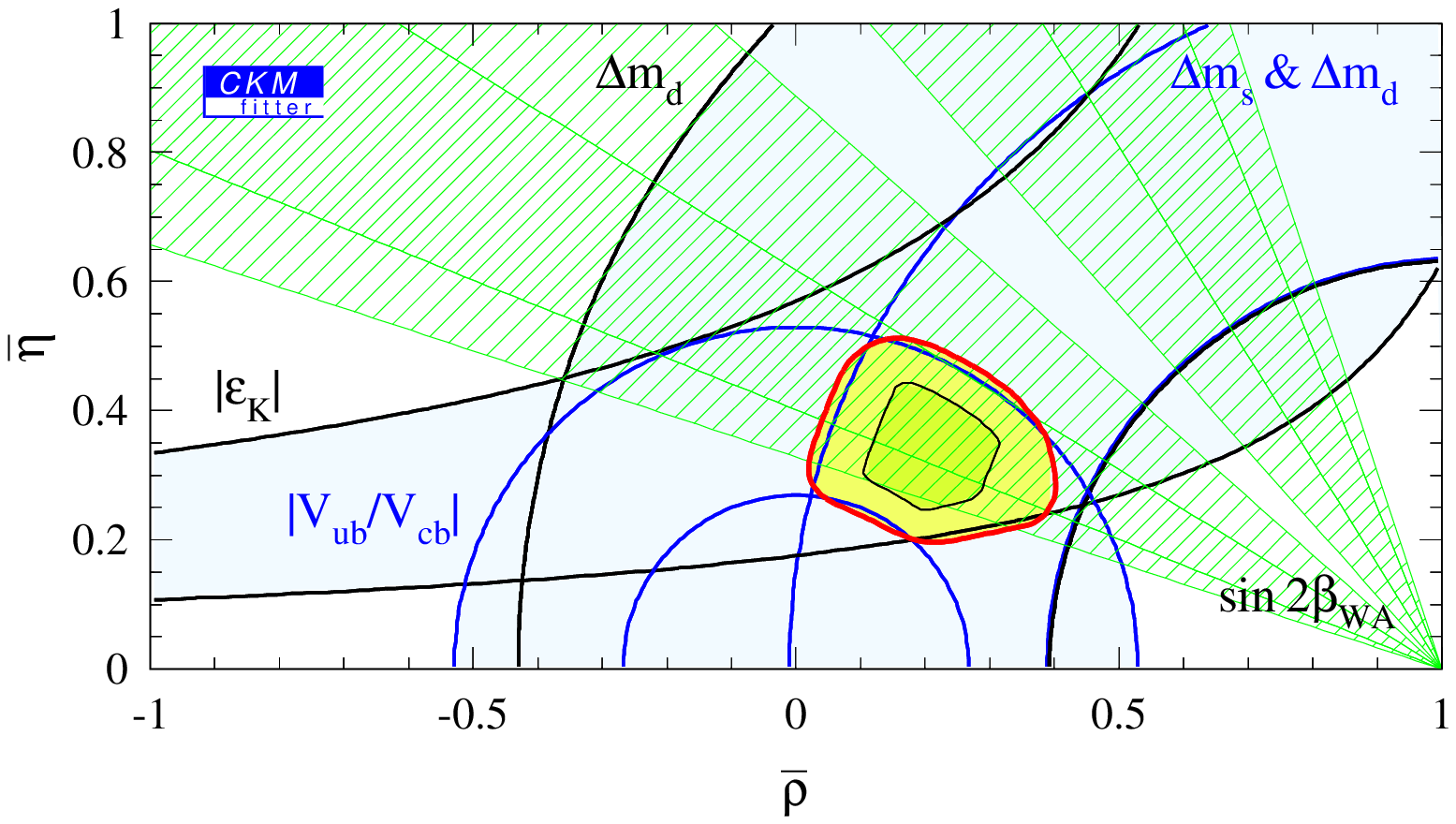}
\caption{Global fit of the unitarity triangle (darker shaded region),
without $\sin 2\beta$ from CP violation in $B\to J/\Psi K_S$.
The constraint from the world average $\sin 2\beta$ 
(hatched areas) is overlaid (H\"ocker et al. 2001).}
\label{fig:utexp}
\end{figure}

$B$ -- $\bar B$ mixing, CP violation in $B$ decays and other loop-induced
reactions of $b$-flavoured hadrons are of great interest and continue
to be pursued by numerous experiments. A central target is the
Cabibbo-Kobayashi-Maskawa (CKM) matrix
\begin{equation}
\label{vckm}
V=\left(\begin{array}{ccc}
V_{ud} & V_{us} & V_{ub}\\
V_{cd} & V_{cs} & V_{cb}\\
V_{td} & V_{ts} & V_{tb}\end{array}\right)\simeq
\left(\begin{array}{ccc}
1-\frac{\lambda^2}{2} & {\lambda} & 
A{\lambda^3}(\varrho-{i\eta})\\
-{\lambda} & 1-\frac{\lambda^2}{2} & A{\lambda^2}\\
A{\lambda^3}(1-\varrho-{i\eta}) & -A{\lambda^2} & 1
\end{array}\right)
\end{equation}
which parametrizes charged-current weak interactions in the
standard model. The second equality in (\ref{vckm}) introduces
the approximate form of the Wolfenstein parametrization, where
the four independent CKM quantities are $\lambda$, $A$, $\varrho$ and
$\eta$. The unitarity triangle, as defined in terms of $\varrho$ and
$\eta$ is shown in Fig. \ref{fig:utdef},
indicating the CP violating angles $\alpha$, $\beta$ and $\gamma$.
The status of the unitarity triangle 
(in terms of $\bar\varrho=\varrho(1-\lambda^2/2)$ and
$\bar\eta=\eta(1-\lambda^2/2)$)
is displayed in Fig. \ref{fig:utexp},
with input from CP violation in the kaon sector ($\varepsilon_K$)
and from $B$ decays ($|V_{ub}/V_{cb}|$, $\Delta m_{B_d}$,
$\Delta m_{B_s}$ and $\sin 2\beta$).

\subsection{$B$ decays -- overview}

A vast number of different $B$-decay observables is available
to further test the standard scenario of flavour dynamics.
One may distinguish the following broad classes.
\begin{itemize}
\item
dominant decays\\
$b\to c\bar ud$, $b\to c\bar cs$, $b\to c l\bar\nu$\\
$\bar B\to D\pi$, $\bar B\to\Psi K$, $\bar B\to D^{(*)} l\bar\nu$
\item
rare decays\\
$b\to u\bar ud$, $b\to u\bar us$, $b\to u l\bar\nu$\\
$\bar B\to \pi\pi$, $\bar B\to\pi K$, $\bar B\to \pi l\bar\nu$, 
$\bar B\to l\bar\nu$
\item
rare and radiative (loop induced) decays\\
$b\to s(d)\gamma$, $b\to s(d) l^+l^-$, $b\to s(d) \nu\bar\nu$\\
$\bar B\to K^{(*)}\gamma$, $\bar B\to\rho\gamma$, $\ldots$
\item
$\Delta B=2$ oscillations\\
$B_d$ -- $\bar B_d$, $B_s$ -- $\bar B_s$ mixing
\end{itemize}

Other obvious classifications are between inclusive and exclusive
processes or between hadronic and (semi)leptonic decays.

A few key properties of $b$-hadrons enhance considerably the
possibilities in $B$ physics both experimentally and theoretically.
First, the smallness of $V_{cb}=0.04$ leads to a long lifetime
of $\tau_B\approx 1.6$ ps. In addition the $b$-quark mass is large
compared with the QCD scale
\begin{equation}
\label{mblqcd}
m_b\gg \Lambda_{QCD}\approx 0.3 {\rm GeV}
\end{equation}
The exact value of $m_b$ depends on the definition.
In particular, the running $\overline{MS}$ mass
$\bar m_b(\bar m_b)=4.2\pm 0.1 {\rm GeV}$,
the pole mass
$m_{b,pole}\approx 4.8 {\rm GeV}$,
whereas the mass of the lightest $b$-hadron is $m_B=5.28 {\rm GeV}$.
The smallness of $\Lambda_{QCD}/m_b$ provides us with a useful
expansion parameter. Together with the property of asymptotic freedom
of QCD and $\alpha_s(m_b)\ll 1$, this opens the possibility of
systematic approximations, which are exploited in the various
applications of heavy quark theory.

\section{Effective weak Hamiltonians}

\begin{figure}
[htbp]
\centering
\includegraphics[width=11.5cm,height=3.5cm]{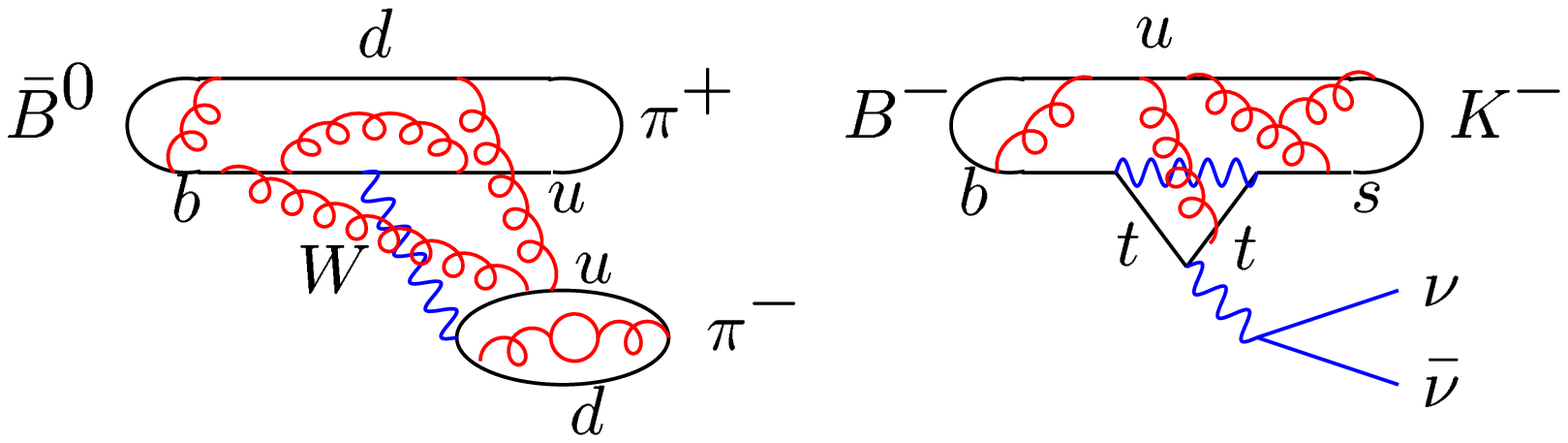}
\caption{QCD effects in weak decays.}
\label{fig:bppbpnn}
\end{figure}

The task of computing weak decays of hadrons represents a complicated
problem in quantum field theory.
Two typical cases, the first-order nonleptonic process
$\bar B^0\to\pi^+\pi^-$, and the loop-induced, second-order weak
transition $B^-\to K^-\nu\bar\nu$ are illustrated in 
Fig. \ref{fig:bppbpnn}.
The dynamics of the decays is determined by a nontrivial interplay
of strong and electroweak forces, which is characterized by several
energy scales of very different magnitude, the $W$ mass, the various
quark masses and the QCD scale:
$m_t$, $M_W\gg$ $m_b$, $m_c\gg$ $\Lambda_{QCD} \gg$ $m_u$, $m_d$, $(m_s)$.
While it is usually sufficient to treat electroweak interactions
to lowest nonvanishing order in perturbation theory, it is necessary
to consider all orders in QCD. Asymptotic freedom still allows us to
compute the effect of strong interactions at short distances
perturbatively. However, since the participating hadrons are bound states 
with light quarks,
confined inside the hadron by long-distance dynamics, it is clear that 
also nonperturbative QCD interactions enter the decay process in an
essential way.

To deal with this situation, we need a method to disentangle long-
and short-distance contributions to the decay amplitude in a systematic
fashion. A basic tool for this purpose is provided by the operator product
expansion (OPE).

\subsection{Operator product expansion}

\begin{figure}
[htbp]
\centering
\includegraphics[width=5cm,height=3.0cm]{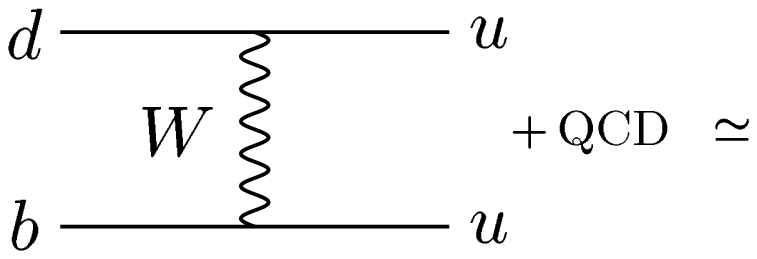}
\raisebox{1.3cm}{{\Large $C\left(\frac{M_W}{\mu}, \alpha_s\right)\cdot$}}
\includegraphics[width=4cm,height=3.0cm]{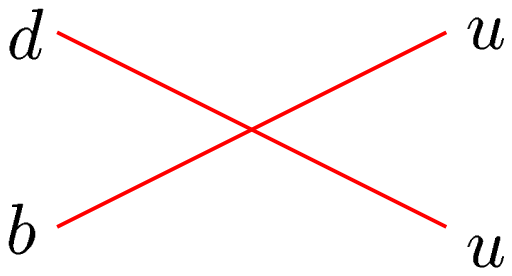}
\caption{OPE for weak decays.}
\label{fig:ope}
\end{figure}

We will now discuss the basic concepts of the OPE for
$B$ meson decay amplitudes. These concepts are of crucial importance
for the theory of weak decay processes, not only in the case
of $B$ mesons, but also
for kaons, mesons with charm, light or heavy baryons and 
weakly decaying hadrons in general.
Consider, for instance, the basic $W$-boson exchange process shown on 
the left-hand side of Fig. \ref{fig:ope}. 
This diagram mediates the
decay of a $b$ quark and triggers the nonleptonic decay of a $B$ meson
such as $\bar B^0\to\pi^+\pi^-$. The quark-level transition shown is
understood to be dressed with QCD interactions of all kinds, including
the binding of the quarks into the mesons. To simplify this
problem, we may look for a suitable expansion parameter, as we are
used to do in theoretical physics. Here, a key feature is provided
by the fact that the $W$ mass $M_W$ is very much heavier than the
other momentum scales $p$ in the problem 
($m_b$, $\Lambda_{QCD}$, $m_u$, $m_d$, $m_s$). We can therefore expand the
full amplitude $A$, schematically, as follows
\begin{equation}\label{acq}
A=C\left(\frac{M_W}{\mu},\alpha_s\right)\cdot\langle Q\rangle
+{\cal O}\left(\frac{p^2}{M^2_W}\right)
\end{equation}
which is sketched in Fig. \ref{fig:ope}. Up to negligible
power corrections of ${\cal O}(p^2/M^2_W)$, the full amplitude
on the left-hand side is written as the matrix element of a local
four-quark operator $Q$, multiplied by a Wilson coefficient $C$.
This expansion in $1/M_W$ is called a (short-distance) operator
product expansion because the nonlocal product of two bilinear
quark-current operators $(\bar du)$ and $(\bar ub)$ that interact
via $W$ exchange, is expanded into a series of local operators.
Physically, the expansion in Fig. \ref{fig:ope} means that the
exchange of the very heavy $W$ boson can be approximated by a
point-like four-quark interaction. With this picture the formal
terminology of the OPE can be expressed in a more intuitive language
by interpreting the local four-quark operator as a four-quark
interaction vertex and the Wilson coefficient as the corresponding
coupling constant. Together they define an effective Hamiltonian
${\cal H}_{eff}=C\cdot Q$, describing weak interactions of light
quarks at low energies.
Ignoring QCD the OPE reads explicitly (in momentum space)
\begin{eqnarray}\label{atree}
A &=& \frac{g^2_W}{8}V^*_{ud}V_{ub}\frac{i}{k^2-M^2_W}
(\bar du)_{V-A}(\bar ub)_{V-A} \nonumber \\
&=& -i\frac{G_F}{\sqrt{2}}V^*_{ud} V_{ub} C\cdot\langle Q\rangle
+{\cal O}\left(\frac{k^2}{M^2_W}\right)
\end{eqnarray}
with $C=1$, $Q=(\bar du)_{V-A}(\bar ub)_{V-A}$ and
\begin{equation}\label{htree}
{\cal H}_{eff}=\frac{G_F}{\sqrt{2}}V^*_{ud} V_{ub}
(\bar du)_{V-A}(\bar ub)_{V-A}
\end{equation}

As we will demonstrate in more detail below after including QCD
effects, the most important property of the OPE in (\ref{acq})
is the {\it factorization \/} of long- and short-distance contributions:
All effects of QCD interactions above some factorization scale $\mu$
(short distances) are contained in the Wilson coefficient $C$.
All the low-energy contributions below $\mu$ (long distances) are
collected into the matrix elements of local operators $\langle Q\rangle$.
In this way the short-distance part of the amplitude can be 
systematically extracted and calculated in perturbation theory.
The problem to evaluate the matrix elements of local operators between
hadron states remains. This task requires in general nonperturbative
techniques, as for example lattice QCD or QCD sum rules, but it is 
considerably simpler than the original problem of the full standard-model 
amplitude.
In some cases also the approximate flavour symmetries of QCD
(isospin, $SU(3)$) can help to determine the nonperturbative input. 
This is true in general for hadronic weak decays.
A decisive advantage of {\it heavy\/} hadrons is the fact that
the heavy-quark mass itself is still large in comparison to
$\Lambda_{QCD}$. The limit $\Lambda_{QCD}/m_b\ll 1$ can then be
exploited, which is achieved, depending on the application,
by using heavy quark effective theory, heavy quark expansion
or QCD factorization for exclusive nonleptonic decays.

The short-distance OPE that we have described, the resulting
effective Hamiltonian, and the factorization property are fundamental
for the theory of $B$ decays. However, the concept of factorization
of long- and short-distance contributions reaches far beyond these
applications. In fact, the idea of factorization, in various forms
and generalizations, is the key to essentially all applications
of perturbative QCD, including the important areas of deep-inelastic
scattering and jet or lepton pair
production in hadron-hadron collisions. The reason is the same in
all cases: Perturbative QCD is a theory of quarks and gluons, but those
never appear in isolation and are always bound inside hadrons.
Nonperturbative dynamics is therefore always relevant to some extent
in hadronic reactions, even if these occur at very high energy or
with a large intrinsic mass scale. Thus, before perturbation theory can be
applied, nonperturbative input has to be isolated in a systematic
way, and this is achieved by establishing the property of 
factorization.
It turns out that the weak effective Hamiltonian for nonleptonic
$B$ decays provides a nice example to demonstrate the general idea of
factorization in simple and explicit terms.

\begin{figure}
[htbp]
\centering
\includegraphics[width=11.5cm,height=3.5cm]{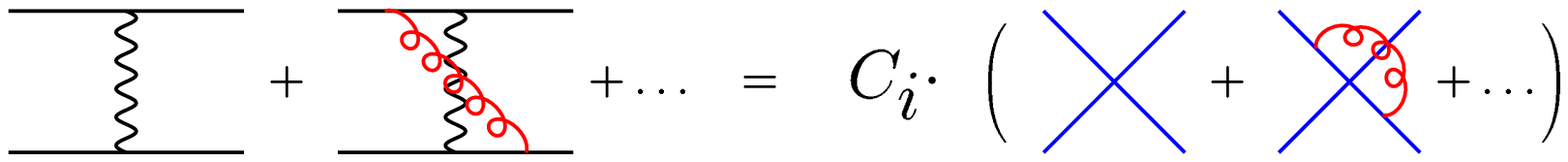}
\vspace*{-2cm}
\caption{Calculation of Wilson coefficients of the OPE.}
\label{fig:opeqcd}
\end{figure}

We would therefore like to discuss the OPE for $B$ decays in more detail, 
including the effects
of QCD, and illustrate the calculation of the Wilson coefficients.
A diagrammatic representation for the OPE is shown in 
Fig. \ref{fig:opeqcd}.
The key to calculating the coefficients $C_i$ is again the
property of factorization. Since factorization implies the separation
of all long-distance sensitive features of the amplitude into the
matrix elements of $\langle Q_i\rangle$, the short-distance quantities
$C_i$ are, in particular, independent of the external states.
This means that the $C_i$ are always the same, no matter whether we
consider the actual physical amplitude where the quarks are bound inside
mesons, or any other, unphysical amplitude with on-shell or even
off-shell external quark lines. Thus, even though we are ultimately
interested in, e.g., $B\to\pi\pi$ amplitudes, for the perturbative
evaluation of $C_i$ we are free to choose any treatment of the
external quarks according to our calculational convenience.
A convenient choice that we will use below is to take all external
quarks massless and with the same off-shell momentum $p$ ($p^2\not= 0$).

The computation of the $C_i$ in perturbation theory then proceeds in
the following steps:

\begin{itemize}
\item 
Compute the amplitude $A$ in the full theory (with $W$ propagator)
for arbitrary external states.
\item
Compute the matrix elements $\langle Q_i\rangle$ with the same
treatment of external states.
\item
Extract the $C_i$ from $A=C_i\, \langle Q_i\rangle$.
\end{itemize}

\begin{figure}
[htbp]
\centering
\hspace*{2cm}
\includegraphics[width=7cm,height=5cm]{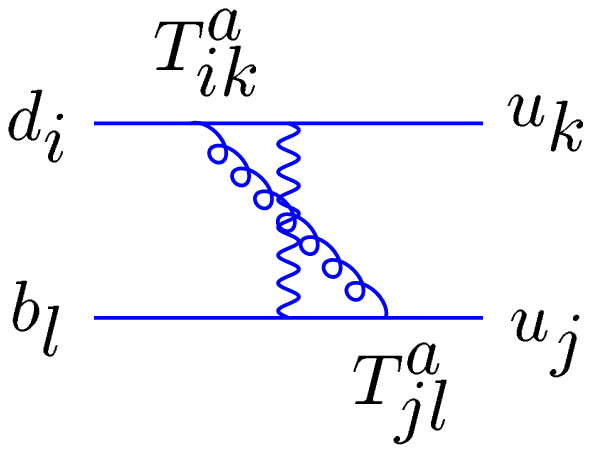}
\vspace*{-2cm}
\caption{QCD correction with colour assignment.}
\label{fig:q1col}
\end{figure}

We remark that with the off-shell momenta $p$ for the quark lines
the amplitude is even gauge dependent and clearly unphysical.
However, this dependence is identical for $A$ and $\langle Q_i\rangle$
and drops out in the coefficients. The actual calculation is most
easily performed in Feynman gauge.
To ${\cal O}(\alpha_s)$ there are four relevant diagrams, the one
shown in Fig. \ref{fig:opeqcd} together with the remaining three
possibilities to connect the two quark lines with a gluon.
Gluon corrections to either of these quark currents need not be
considered, they are the same on both sides of the OPE and drop out
in the $C_i$.
The operators that appear on the right-hand side follow from the
actual calculations.
Without QCD corrections there is only one operator of dimension 6
\begin{equation}\label{q2def}
Q_1=(\bar d_i u_i)_{V-A}(\bar u_j b_j)_{V-A}
\end{equation}
where the colour indices have been made explicit.
To ${\cal O}(\alpha_s)$ QCD generates another operator
\begin{equation}\label{q1def}
Q_2=(\bar d_i u_j)_{V-A}(\bar u_j b_i)_{V-A}
\end{equation}
which has the same Dirac and flavour structure, but a different
colour form.
Its origin is illustrated in Fig. \ref{fig:q1col},
where we recall the useful identity for $SU(N)$ Gell-Mann matrices
\begin{equation}\label{tata}
(\bar d_i T^a_{ik} u_k) (\bar u_j T^a_{jl} b_l)=
-\frac{1}{2N}(\bar d_i u_i) (\bar u_j b_j)
+\frac{1}{2}(\bar d_i u_j) (\bar u_j b_i)
\end{equation}
It is convenient to employ a different operator basis, defining
\begin{equation}\label{qpm}
Q_\pm=\frac{Q_1\pm Q_2}{2}
\end{equation}
The corresponding coefficients are then given by
\begin{equation}\label{cpm1}
C_\pm=C_1\pm C_2
\end{equation}
If we denote by $S_\pm$ the spinor expressions that correspond
to the operators $Q_\pm$ (in other words: the tree-level matrix
elements of $Q_\pm$), the full amplitude can be written as
\begin{equation}\label{aft}
A=\left(1+\gamma_+ \alpha_s \ln\frac{M^2_W}{-p^2}\right)S_+
+ \left(1+\gamma_- \alpha_s \ln\frac{M^2_W}{-p^2}\right)S_-
\end{equation}
Here we have focused on the logarithmic terms and dropped
a constant contribution (of order $\alpha_s$, but nonlogarithmic).
Further, $p^2$ is the virtuality of the quarks and $\gamma_\pm$
are numbers that we will specify later on.
We next compute the matrix elements of the operators in the effective
theory, using the same approximations, and find
\begin{equation}\label{qme}
\langle Q_\pm\rangle=
\left(1+\gamma_\pm \alpha_s\left(\frac{1}{\varepsilon}+
\ln\frac{\mu^2}{-p^2}\right)\right) S_\pm
\end{equation}
The divergence that appears in this case has been regulated in
dimensional regularization ($D=4-2\varepsilon$ dimensions).
Requiring
\begin{equation}\label{acqpm}
A=C_+ \langle Q_+\rangle + C_- \langle Q_-\rangle
\end{equation}
we obtain
\begin{equation}\label{cll}
C_\pm = 1+\gamma_\pm \alpha_s \ln\frac{M^2_W}{\mu^2}
\end{equation}
where the divergence has been subtracted in the minimal subtraction
scheme.
The effective Hamiltonian we have been looking for then reads
\begin{equation}\label{hcqpm}
{\cal H}_{eff}=\frac{G_F}{\sqrt{2}}V^*_{ud}V_{ub}
\left( C_+(\mu) Q_+ + C_-(\mu) Q_-\right)
\end{equation}
with the coefficients $C_\pm$ determined in (\ref{cll}) to
${\cal O}(\alpha_s\,{\rm log})$ in perturbation theory.
The following points are worth noting:

\begin{itemize}
\item
The $1/\varepsilon$ (ultraviolet) divergence in the effective theory
(\ref{qme}) reflects the $M_W\to\infty$ limit. This can be seen from
the amplitude in the full theory (\ref{aft}),
which is finite, but develops a logarithmic singularity in this limit.
Consequently, the renormalization in the effective theory is directly
linked to the $\ln M_W$ dependence of the decay amplitude.
\item
We observe that although $A$ and $\langle Q_\pm\rangle$ both
depend on the long-distance properties of the external states
(through $p^2$), this 
de\-pen\-dence has drop\-ped out in $C_\pm$.
Here we see explicitly how factorization is realized.
Technically, to ${\cal O}(\alpha_s\,{\rm log})$, factorization is
equivalent to splitting the logarithm of the full amplitude 
according to
\begin{equation}
\ln\frac{M^2_W}{-p^2}=\ln\frac{M^2_W}{\mu^2}+\ln\frac{\mu^2}{-p^2}
\end{equation}
Ultimately the logarithms stem from loop momentum integrations
and the range of large momenta, between
$M_W$ and the factorization scale $\mu$, is indeed separated into the
Wilson coefficients.
\item
To obtain a decay amplitude from ${\cal H}_{eff}$ in
(\ref{hcqpm}), the matrix elements
$\langle f|Q_\pm|\bar B\rangle(\mu)$ have to be taken, normalized at a
scale $\mu$. An appropriate value for $\mu$ is close to the $b$-quark mass
scale in order not to introduce an unnaturally large scale into
the calculation of $\langle Q\rangle$. 
\item
The factorization scale $\mu$ is unphysical. It cancels between
Wilson coefficient and hadronic matrix element, to a given order
in $\alpha_s$, to yield a scale independent decay amplitude.
The mechanism of this cancellation to ${\cal O}(\alpha_s)$ is
clear from the above example (\ref{aft}) -- (\ref{cll}).
\item
In the construction of ${\cal H}_{eff}$ the $W$-boson is said to
be ``integrated out'', that is, removed from the effective theory as
an explicit degree of freedom. Its effect is still implicitly
contained in the Wilson coefficients. The extraction of these
coeffcients is often called a ``matching calculation'',
matching the full to the effective theory by ``adjusting'' the
couplings $C_\pm$.
\item
If we go beyond the leading logarithmic approximation
${\cal O}(\alpha_s\log)$ and include the finite corrections
of ${\cal O}(\alpha_s)$ in (\ref{aft}), (\ref{qme}), an ambiguity
arises when renormalizing the divergence in (\ref{qme}) (or,
equivalently, in the Wilson coefficients $C_\pm$).
This ambiguity consists in what part of the full 
(non-logarithmic) ${\cal O}(\alpha_s)$ term is attributed to the matrix
elements, and what part to the Wilson coefficients. In other words,
coefficients and matrix elements become {\it scheme dependent},
that is, dependent on the renormalization scheme, beyond the leading
logarithmic approximation. The scheme dependence is 
unphysical and cancels in the product of coefficients and matrix
elements. Of course, both quantities have to be evaluated in the
same scheme to obtain a consistent result. The renormalization scheme
is determined in particular by the subtraction constants
(minimal or non-minimal subtraction of $1/\varepsilon$ poles), and
also by the definition of $\gamma_5$ used in $D\not= 4$ dimensions
in the context of dimensional regularization. 
\item
Finally, the effective Hamiltonian (\ref{hcqpm}) can be considered
as a modern version of the old Fermi theory for weak interactions.  
It is a systematic 
low-energy approximation to the standard model for $b$-hadron decays
and provides the basis for any further analysis.
\end{itemize}

\subsection{Renormalization group}

Let us have a closer look at the Wilson coefficents, which read
explicitly
\begin{equation}\label{cpmg}
C_\pm=1+\frac{\alpha_s(\mu)}{4\pi}\frac{\gamma^{(0)}_\pm}{2}
\ln\frac{\mu^2}{M^2_W}
\qquad
\gamma^{(0)}_\pm =\left\{
\begin{array}{c}
4\\ -8
\end{array}\right.
\end{equation}
where we have now specified the exact form of the
${\cal O}(\alpha_s\,{\rm log})$ correction.
Numerically the factor $\alpha_s(m_b) \gamma^{(0)}_\pm/(8\pi)$
is about $+3.5\%$ ($-7\%$), a reasonable size for a perturbative
correction (we used $\alpha_s(\mu=4.2\,{\rm GeV})=0.22$).
However, this term comes with a large logarithmic factor of
$\ln(\mu^2/M^2_W)=-6$, for an appropriate scale of $\mu=4.2\,{\rm GeV}$.
The total correction to $C_\pm=1$ in (\ref{cpmg}) is then
$-21\%$ ($42\%$)! The presence of the large logarithm
spoils the validity of a straightforward perturbative expansion,
despite the fact that the coupling constant itself is still
reasonably small. This situation is quite common in renormalizable
quantum field theories. Logarithms appear naturally and can become
very large when the problem involves very different scales.
The general situation is indicated in the following table, where we 
display the form of the correction terms in higher orders, denoting
$\ell\equiv\ln(\mu/M_W)$
\begin{equation}\nonumber
\begin{array}{cccc}
{\rm LL} & {\rm NLL} & & \\
\alpha_s\ell & \alpha_s & & \\
\alpha^2_s\ell^2 & \alpha^2_s\ell & \alpha^2_s & \\
\alpha^3_s\ell^3 & \alpha^3_s\ell^2 & \alpha^3_s\ell & \alpha^3_s \\
\downarrow & \downarrow & & \\
{\cal O}(1) & {\cal O}(\alpha_s) & & \\
\end{array}
\end{equation}
In ordinary perturbation theory the expansion is organized according
to powers of $\alpha_s$ alone, corresponding to the rows in the
above scheme. This approach is invalidated by the large logarithms
since $\alpha_s\ell$, in contrast to $\alpha_s$, is no longer a
small parameter, but a quantity of order 1.
The problem can be resolved by resumming the terms 
$(\alpha_s\ell)^n$ to all orders $n$.
The expansion is then reorganized in terms of columns of the
above table. The first column is of ${\cal O}(1)$ and yields
the leading logarithmic approximation, the second column gives a
correction of relative order $\alpha_s$, and so forth. 
Technically the reorganization is achieved by solving the
renormalization group equation (RGE) for the Wilson coefficients.
The RGE is a differential equation describing the change of 
$C_\pm(\mu)$ under a change of scale. To leading order this equation
can be read off from (\ref{cpmg})
\begin{equation}\label{rgecpm}
\frac{d}{d\ln\mu} C_\pm(\mu)=
\frac{\alpha_s}{4\pi}\gamma^{(0)}_\pm\cdot C_\pm(\mu)
\end{equation}
$(\alpha_s/4\pi)\gamma^{(0)}_\pm$ are called the anomalous dimensions
of $C_\pm$. To understand the term ``dimension'', compare with the
following relation for the quantity $\mu^n$, which has (energy)
dimension $n$:
\begin{equation}\label{dmun}
\frac{d}{d\ln\mu}\mu^n = n\cdot \mu^n
\end{equation}
The analogy is obvious. 
Of course, the $C_\pm(\mu)$ are dimensionless numbers in the
usual sense; they can depend on the energy scale $\mu$ only because
there is another scale, $M_W$, present under the logarithm in 
(\ref{cpmg}). Their ``dimension'' is therefore more precisely
called a scaling dimension, measuring the rate of change of $C_\pm$
with a changing scale $\mu$. The nontrivial scaling dimension
derives from ${\cal O}(\alpha_s)$ loop corrections and is thus a 
genuine quantum effect. Classically the coefficients are 
scale invariant, $C_\pm\equiv 1$. Whenever a symmetry that holds
at the classical level is broken by quantum effects, we speak of an
``anomaly''. Hence, the $\gamma^{(0)}_\pm$ represent the
anomalous (scaling) dimensions of the Wilson coefficients.

We can solve (\ref{rgecpm}), using
\begin{equation}\label{asb0c}
\frac{d\alpha_s}{d\ln\mu}=-2\beta_0\frac{\alpha^2_s}{4\pi}
\quad
\beta_0=\frac{33-2f}{3}\quad C_\pm(M_W)=1
\end{equation}
and find
\begin{equation}\label{cpmll}
C_\pm(\mu)=
\left[\frac{\alpha_s(M_W)}{\alpha_s(\mu)}
  \right]^{\frac{\gamma^{(0)}_\pm}{2\beta_0}}=
\left[\frac{1}{1+\beta_0\frac{\alpha_s(\mu)}{4\pi}\ln\frac{M^2_W}{\mu^2}}
\right]^{\frac{\gamma^{(0)}_\pm}{2\beta_0}}
\end{equation}
This is the solution for the Wilson coefficients $C_\pm$ in leading
logarithmic approximation, that is to leading order in RG improved
perturbation theory. The all-orders resummation of $\alpha_s\,{\rm log}$
terms is apparent in the final expression in (\ref{cpmll}).

\subsection{$\Delta B=1$ effective Hamiltonian}

\begin{figure}
[htbp]
\centering
\hspace*{4.2cm}
\includegraphics[width=7cm,height=5cm]{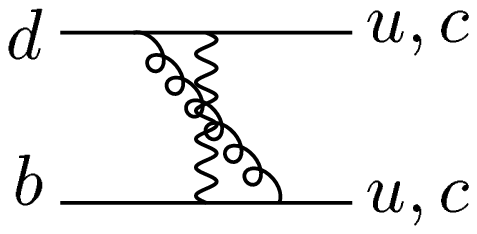}
\vspace*{-3cm}
\caption{QCD correction to $W$ exchange.}
\label{fig:b1cc}
\end{figure}

\begin{figure}
[htbp]
\centering
\hspace*{4.2cm}
\includegraphics[width=7cm,height=5cm]{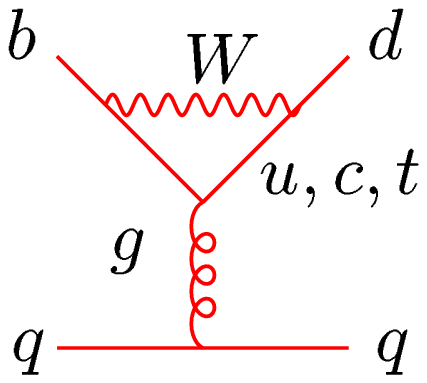}
\vspace*{-2.5cm}
\caption{QCD-penguin diagram.}
\label{fig:b1peng}
\end{figure}

In this section we will complete the discussion of the
$\Delta B=1$ effective Hamiltonian.
So far we have considered the operators
\begin{eqnarray}\label{q12def}
Q^p_1 &=& (\bar d_i p_i)_{V-A}(\bar p_j b_j)_{V-A} \\
Q^p_2 &=& (\bar d_i p_j)_{V-A}(\bar p_j b_i)_{V-A} 
\end{eqnarray}
which come from the simple $W$-exchange graph and the corresponding
QCD corrections (Fig. \ref{fig:b1cc}). We have slightly
generalized our previous notation, allowing for the cases $p=u$, $c$.
In addition, there is a further type of diagram at ${\cal O}(\alpha_s)$,
which we have omitted until now: the QCD-penguin diagram shown
in Fig. \ref{fig:b1peng}. 
It gives rise to the four new operators
\begin{eqnarray}\label{q36def}
Q_3 &=& (\bar d_i b_i)_{V-A}\sum_q(\bar q_j q_j)_{V-A} \\
Q_4 &=& (\bar d_i b_j)_{V-A}\sum_q(\bar q_j q_i)_{V-A} \\
Q_5 &=& (\bar d_i b_i)_{V-A}\sum_q(\bar q_j q_j)_{V+A} \\
Q_6 &=& (\bar d_i b_j)_{V-A}\sum_q(\bar q_j q_i)_{V+A} 
\end{eqnarray}
Two structures appear when the light-quark current $(\bar qq)_V$
from the bottom end of the diagram is split into $V-A$ and $V+A$
parts. In turn, each of those comes in two colour forms in a way
similar to $Q_1$ and $Q_2$.
Finally, one further gauge-invariant operator of dimension
six appears in the matching procedure, the
chromomagnetic penguin operator 
\begin{equation}\label{q8def}
Q_{8g} = -\frac{g}{8\pi^2}m_b\, 
        \bar d_i\sigma^{\mu\nu}(1+\gamma_5)T^a_{ij} b_j\, G^a_{\mu\nu}
\end{equation}
This operator corresponds to the diagrams in Fig. \ref{fig:b1peng}
with the lower quark line omitted. The gluon is thus an external
field, represented in (\ref{q8def})
by the field-strength tensor $G^a_{\mu\nu}$.
Note that the characteristic tensor current necessitates a
helicity flip in the $b\to d$ transition, which is accompanied
by a factor of the quark mass $m_b$ (the effect of $m_d$ is
neglected). The contribution of $Q_{8g}$ would be very small
for the Hamiltonian of $K$ decays, which only involves light
external quarks, but it is unsuppressed for $b$ decays.

The operators $Q_1,\ldots ,Q_6$, $Q_{8g}$ mix under renormalization, that is
the RGE for their Wilson coefficients is governed by a matrix
of anomalous dimensions, generalizing (\ref{rgecpm}).
In this way the RG evolution of $C_{1,2}$ affects the evolution
of $C_3,\ldots , C_6$, $C_{8g}$. On the other hand $C_{1,2}$ remain unchanged
in the presence of the penguin operators $Q_3,\ldots ,Q_6$, $Q_{8g}$,
so that the results for $C_{1,2}$ derived above are still valid.

The construction of the effective Hamiltonian follows the principles
we have discussed in the previous sections.
First the Wilson coefficients $C_i(\mu_W)$, $i=1,\ldots, 6, 8g$,
are determined at a large scale $\mu_W={\cal O}(M_W,m_t)$ to a given
order in perturbation theory. In this step both the $W$ boson and the
heavy top quark are integrated out. Since the renormalization scale
is chosen to be $\mu_W={\cal O}(M_W,m_t)$, no large logarithms
appear and straightforward perturbation theory can be used for the
matching calculation.
The anomalous dimensions are computed from the divergent parts of
the operator matrix elements, which correspond to the 
UV-renormalization of the Wilson coefficients. Solving the
RGE the $C_i$ are evolved from $\mu_W$ to a scale 
$\mu={\cal O}(m_b)$ in a theory with $f=5$ active flavours
$q=u,d,s,c,b$. 
The terms taken into account in the RG improved perturbative
evaluation of $C_i(\mu)$ are, schematically:

\begin{center}
LO: $\left(\alpha_s\ln\frac{M_W}{\mu}\right)^n$,
\hspace*{2cm}
NLO: $\alpha_s\left(\alpha_s\ln\frac{M_W}{\mu}\right)^n$,
\end{center}

\noindent
at leading and next-to-leading order, respectively.

The final result for the $\Delta B=1$ effective Hamiltonian
can be written as
\begin{equation}\label{hdb1}
{\cal H}^{\Delta B=1}_{eff}=\frac{G_F}{\sqrt{2}}\sum_{p=u,c} \lambda_p
\left[ C_1 Q^p_1 + C_2 Q^p_2 +\sum_{i=3,\ldots ,6,{8g}} C_i Q_i\right]\, 
+ \,{\rm h.c.}
\end{equation}
where $\lambda_p\equiv V^*_{pd}V_{pb}$.
In principle there are three different CKM factors,
$\lambda_u$, $\lambda_c$ and $\lambda_t$, corresponding to the
different flavours of up-type quarks that can participate
in the charged-current weak interaction. Using CKM unitarity, one
of them can be eliminated. If we eliminate $\lambda_t$,
we arrive at the CKM structure of (\ref{hdb1}).

The Hamiltonian in (\ref{hdb1}) is the basis for computing
nonleptonic $b$-hadron decays within the standard model
(to lowest order in electroweak interactions)
with $\Delta B=1$ and $\Delta S$, $\Delta C=0$.
The Hamiltonian for $b$-decays with different flavour
quantum numbers of the light quarks has a completely
analogous form. For instance $\Delta B=1$ transitions
with a simultaneous change in strangeness, $\Delta S=1$,
are simply described by  (\ref{hdb1}) after the replacement
$d\to s$.
When new physics is present at some higher energy scale,
the effective Hamiltonian can be derived in an analogous way.
The matching calculation at the high scale $\mu_W$ will give
new contributions to the coefficients $C_i(\mu_W)$, the initial
conditions for the RG evolution. In general, new operators may
also be induced.
The Wilson coefficients $C_i$ are known in the standard
model at NLO. A more detailed account of 
${\cal H}^{\Delta B=1}_{eff}$ and information on the technical
aspects of the necessary calculations can be found in
(Buchalla, Buras \& Lautenbacher 1996) and (Buras 1998).

\subsection{$B$ -- $\bar B$ mixing at NLO}

\begin{figure}
[htbp]
\centering
\includegraphics[width=6cm,height=3.6cm]{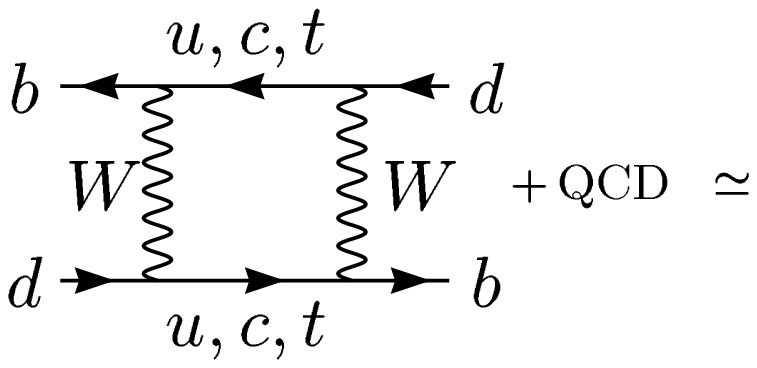}
\raisebox{1.6cm}{{\Large $C(\mu_t)\cdot$}}
\raisebox{0.4cm}{\includegraphics[width=4cm,height=3.0cm]{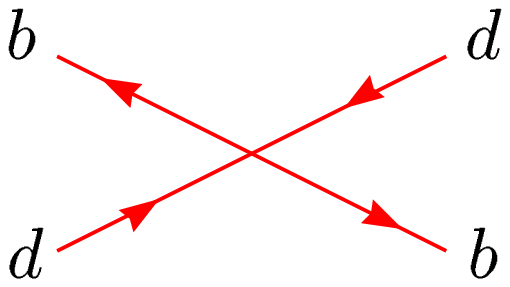}}
\caption{OPE for $B$ -- $\bar B$ mixing.}
\label{fig:db2match}
\end{figure}

In the following section we present the effective Hamiltonian
for $\Delta B=2$ transition, which is relevant for
$B$ -- $\bar B$ mixing. In this case only a single operator
contributes. The form of the Hamiltonian is therefore particularly
simple. We use this example to illustrate the structure of 
Wilson coefficients at next-to-leading order.
The mass difference $\Delta m_B$ in the $B$ -- $\bar B$ system is
related to the effective Hamiltonian ${\cal H}^{\Delta B=2}_{eff}$ by
\begin{equation}
\label{dmbdef}
\Delta m_B=2 |M_{12}|_B=
\frac{1}{m_B}\left|\langle\bar B|{\cal H}^{\Delta B=2}_{eff}|B\rangle\right|
\end{equation}
In order to construct ${\cal H}^{\Delta B=2}_{eff}$, the full standard
model amplitude for $\Delta B=2$ transitions is matched onto the
effective theory amplitude at the matching scale
$\mu_t={\cal O}(m_t)={\cal O}(M_W)$. This is sketched in 
Fig. \ref{fig:db2match}.

There is only one local operator
\begin{equation}\label{qdb2}
Q=(\bar bd)_{V-A} (\bar bd)_{V-A}
\end{equation}
The Wilson coefficient, up to next-to-leading order, can be written as
\begin{equation}\label{cmut}
C(\mu_t)=C^{(0)}(\mu_t)+\frac{\alpha_s}{4\pi} C^{(1)}(\mu_t)
\end{equation}
where $C^{(0)}$ is the lowest order result and $C^{(1)}$ comes
from the corrections with one-gluon exchange.
The RG evolution from the high scale $\mu_t$ down to a scale
$\mu={\cal O}(m_b)$ has the form
\begin{equation}\label{cmucmut}
C(\mu)=\left[1+\frac{\alpha_s(\mu)-\alpha_s(\mu_t)}{4\pi}J_5\right]
\cdot\left[\frac{\alpha_s(\mu_t)}{\alpha_s(\mu)}\right]^{6/23}\cdot C(\mu_t)
\end{equation}
The second factor on the right-hand side is familiar from the
leading logarithmic approximation (the only difference is that at
NLO the two-loop expression for $\alpha_s(\mu)$ has to be used).
The first factor represents the next-to-leading order correction.
Here $J_5$ is a scheme-dependent constant, which in the usual,
so-called NDR scheme reads $J_5=5165/3174$.

We now have the ingredients to write the effective Hamiltonian
up to NLO precision
\begin{eqnarray}\label{heffdb2}
{\cal H}^{\Delta B=2}_{eff} &=& 
{\frac{G^2_F M^2_W}{16\pi^2}(V^*_{tb}V_{td})^2}\cdot C(\mu)\, Q \\
&=& {\frac{G^2_F M^2_W}{16\pi^2}(V^*_{tb}V_{td})^2}
S_0(x_t)\eta_B [\alpha_s(\mu)]^{\frac{-6}{23}}
\left[1+\frac{\alpha_s(\mu)}{4\pi}J_5\right] {Q}\nonumber
\end{eqnarray}
The result is entirely dominated by the top-quark contribution.
It is common practice to separate the coefficient $C(\mu)$ into the
function $S_0(x_t)$ ($x_t=m^2_t/M^2_W$), which would be the coefficient
in the absence of QCD effects, into the terms that depend on 
$\alpha_s(\mu)$, and the remainder, which is defined as the QCD-correction
factor $\eta_B$. This has been done in the second equation in
(\ref{heffdb2}). Taking the matrix element of ${\cal H}^{\Delta B=2}_{eff}$
between the $B$ and the $\bar B$ state and using (\ref{dmbdef}) gives
\begin{equation}\label{dmb}
\Delta m_B =
\frac{G^2_F M^2_W}{6\pi^2} |V^*_{tb}V_{td}|^2
S_0(x_t)\eta_B \, B_B f^2_B m_B
\end{equation}
One encounters the hadronic matrix element of $Q$, which is written as
\begin{equation}\label{qbmu}
\langle\bar B|Q|B\rangle(\mu)\equiv\frac{8}{3}f^2_B m^2_B B_B(\mu)
\end{equation}
defining the (scale and scheme dependent) hadronic parameter $B_B(\mu)$.
The combination
\begin{equation}\label{bbmu}
B_B\equiv B_B(\mu)
[\alpha_s(\mu)]^{\frac{-6}{23}}
\left[1+\frac{\alpha_s(\mu)}{4\pi}J_5\right] 
\end{equation}
is formally scale and scheme independent and has been used in (\ref{dmb}).
The parameter $B_B(\mu)$ is a nonperturbative quantity and has to be
determined e.g. by lattice calculations. At present the value of $B_B$
is still very uncertain, in contrast to the short-distance QCD corrections,
which are precisely known. A numerical illustration is given in 
Table 1, where we have put $B_B(\mu)=0.9$ as an example.
\begin{table}[h]
\label{tab:etabnum}
\caption[]{}
\begin{center}
\vspace*{0.4cm}
\begin{tabular}{|cccc|c|}
\hline
\multicolumn{3}{|c|}{${0.846}$} & ${0.9}$ & ${=0.761}$\\
\hline
${\eta_B}$ & $[\alpha_s(\mu)]^{\frac{-6}{23}}$ &
$\left[1+\frac{\alpha_s(\mu)}{4\pi}J_5\right]$ & ${B_B(\mu)}$ & \\
\hline
${0.551}$ & \multicolumn{3}{|c|}{${1.38}$} & ${=0.761}$\\
\hline
\end{tabular}
\end{center}
\end{table}

Two different definitions of a short-distance QCD factor can be 
considered, depending on where the terms with $\alpha_s(\mu)$ are
included. One possibility is to include them with $\eta_B$ into
a Wilson coefficient (=0.846), which is to be multiplied by the
hadronic matrix element $B_B(\mu)=0.9$. The other possibility is the
formally scheme independent separation into $\eta_B=0.551$ and
$B_B=1.38$ (for $\eta_B$ this is the precise result; $B_B=1.38$ is only
true in our example).
The purpose of this exercise is to remind us that different definitions
are sometimes employed for the parameter $B_B$ and care has to be taken
which one is being used, in order to combine it with the appropriate
short-distance corrections. We can also see that the large deviation
of the QCD correction factor $\eta_B$ from 1 is merely a consequence
of pulling out the large factor $[\alpha_s(\mu)]^{-6/23}$.
It is somewhat artificial and does certainly not indicate a problem
for perturbation theory. In fact, the coefficient 0.846 is the one
that has the proper limit, approaching 1 as $\alpha_s\to 0$.
It is indeed much closer to unity in accordance with the expectation
for a perturbative correction factor. Still, the use of $\eta_B=0.551$
is often adopted due to its formally scheme invariant definition.

An important application is the ratio of the mass differences
for $B_d$ and $B_s$ mesons, for which (\ref{dmb}) implies
\begin{equation}\label{dmbdbs}
\frac{\Delta m_{B_d}}{{\Delta m_{B_s}}}=
\left|\frac{V_{td}}{{V_{ts}}}\right|^2
\frac{m_{B_d}f^2_{B_d} B_{B_d}}{{m_{B_s}f^2_{B_s} B_{B_s}}}
\end{equation}
This quantity is a very useful measure of $|V_{td}/V_{ts}|$.
All other short-distance physics (top-dependence, $\eta_B$) has
dropped out. Hadronic uncertainties are reduced in the ratio
of matrix elements, which is 1 in the limit of unbroken
$SU(3)$ flavour symmetry. The cancellation of the short-distance
contribution is a direct consequence of the factorization property
of the OPE.
Lattice calculations give for the ratio of matrix elements
(H\"ocker et al. 2001, and refs. therein)
\begin{equation}\label{xibdbs}
\frac{f_{B_s} \sqrt{B_{B_s}}}{f_{B_d} \sqrt{B_{B_d}}}=1.16\pm 0.06
\end{equation}
The ratio $\Delta m_{B_d}/\Delta m_{B_s}$ is a very powerful constraint
for the unitarity triangle, as can be seen in Fig. \ref{fig:utexp}.

\section{Heavy quark effective theory}

\subsection{Basic formalism}

Heavy quark effective theory is an effective field theory designed
to systematically exploit the simplifications of QCD interactions
in the heavy-quark limit for the case of hadrons containing a single
heavy quark. The HQET Lagrangian can be derived as follows. We start
from the usual QCD Lagrangian for a heavy-quark field $\Psi$ with
mass $m$
\begin{equation}\label{lqcd}
{\cal L}=\bar\Psi i\not\!\! D \Psi - m \bar\Psi\Psi
\end{equation}
with the covariant derivative
\begin{equation}\label{dcov}
D_\mu=\partial_\mu-i g T^a A^a_\mu
\end{equation}
The heavy-quark momentum can be decomposed as
\begin{equation}\label{pmvk}
p=m v+ k
\end{equation}
where $v$ is the 4-velocity of the heavy {\it hadron}.
Once $m v$, the large kinematical part of the momentum is singled out,
the remaining component $k$ is determined by soft QCD bound state
interactions, and thus $k={\cal O}(\Lambda_{QCD})\ll m$.
We next decompose the quark field $\Psi$ into
\begin{eqnarray}
h_v(x)&\equiv& e^{im v\cdot x}\frac{1+\not\! v}{2}\Psi(x)\label{hvdef}\\
H_v(x)&\equiv& e^{im v\cdot x}\frac{1-\not\! v}{2}\Psi(x)\label{hhvdef}
\end{eqnarray}
which implies
\begin{equation}\label{psihv}
\Psi(x)=e^{-im v\cdot x}\left( h_v(x)+H_v(x)\right)
\end{equation}
The expressions $(1\pm\not\! v)/2$ are projection operators.
Their action represents the covariant generalization of decomposing
$\Psi$ into upper and lower components. 
Using the standard representation for $\gamma$-matrices,
this is evident in the rest frame where $\not\! v=\gamma^0$.
Note also that the equation of motion with respect to the large
momentum components, $m(\not\! v - 1)h_v=0$, is manifest for $h_v$.

The exponential factor $\exp(i m v\cdot x)$ in (\ref{hvdef}),
(\ref{hhvdef}) removes the large-frequency part of the $x$-dependence
in $\Psi(x)$ resulting from the large momentum $m v$. Consequently,
the $x$-dependence of $h_v$, $H_v$ is only governed by the small
residual momentum and derivatives acting on $h_v$ and $H_v$
count as ${\cal O}(\Lambda_{QCD})$.
(Our sign conventions are appropriate for a heavy {\it quark}.
To describe the case of a heavy {\it anti-quark}, similar definitions
are valid with the sign of $v$ reversed.)

Multplying the QCD e.o.m. $(i\not\!\!\! D - m)\Psi=0$ with the
projectors $(1 - \not\!\! v)/2$ and $(1 + \not\! v)/2$, and using 
(\ref{hvdef}) -- (\ref{psihv}) and the definition
\begin{equation}\label{dperp}
D^\mu_\perp\equiv D^\mu- v^\mu v\cdot D
\end{equation}
we obtain the coupled system of equations
\begin{eqnarray}
i v\cdot D h_v &=& -i \not\!\! D_\perp H_v \label{heom1}\\
(i v\cdot D + 2m) H_v &=& i \not\!\! D_\perp h_v \label{heom2}
\end{eqnarray}
They represent the e.o.m. in terms of $h_v$ and $H_v$. The second equation
implies that $H_v={\cal O}(\Lambda_{QCD}/m) h_v$ by power counting.
Hence $H_v$ is suppressed with respect to $h_v$ in the heavy-quark limit.
In other words, $h_v$ contains the large components, $H_v$ the small
components of $\Psi$. 

The HQET Lagrangian is obtained starting from (\ref{lqcd}), expressing
$\Psi$ in terms of $h_v$, $H_v$ and eliminating $H_v$ using (\ref{heom2}).
We find
\begin{equation}\label{lhqet}
{\cal L} = \bar h_v i v\cdot D h_v+
\bar h_v i\not\!\! D_\perp\frac{1}{i v\cdot D+2m}i\not\!\! D_\perp h_v
\end{equation} 
Alternatively, $H_v$ as obtained from (\ref{heom2}) in terms of $h_v$
can be inserted into (\ref{heom1}) to yield the e.o.m. for $h_v$.
This equation is just the e.o.m. implied by (\ref{lhqet})
(upon variation with respect to $\bar h_v$, i.e. 
$\delta{\cal L}/\delta\bar h_v =0$). The Lagrangian may thus 
be written down immediately given the e.o.m. for the field $h_v$.

The second term in (\ref{lhqet}) contains the nonlocal operator
$(i v\cdot D + 2m)^{-1}$. It can be expanded in powers of
$\Lambda_{QCD}/m$ to yield a series of local operators. Keeping
only the leading-power correction we can simply replace
$(i v\cdot D + 2m)^{-1}$ by $(2m)^{-1}$ and get
\begin{equation}\label{lhqet1}
{\cal L} = \bar h_v i v\cdot D h_v+
\frac{1}{2m}\bar h_v(i D_\perp)^2 h_v+
\frac{g}{4m}\bar h_v\sigma^{\mu\nu}G_{\mu\nu}h_v
\end{equation}
Let us discuss some important aspects of this result.
\begin{itemize}
\item
The first term on the r.h.s. of (\ref{lhqet1}) is the basic,
lowest-order Lagrangian of HQET. It describes the ``residual'' QCD
dynamics of the heavy quark once the kinematic dependence on $m$ is 
separated out. Since there is no longer any reference to the mass $m$,
the only parameter to distinguish quark flavours, this term is
flavour symmetric: The dynamics is the same for $b$ and $c$ quarks
in the static limit. Since the operator $v\cdot D$ contains no
$\gamma$-matrices, which would act on the spin degrees of freedom,
the leading HQET Lagrangian also exhibits a spin symmetry.
This corresponds to the decoupling of the heavy-quark spin in the
$m\to\infty$ limit.
Together, we have the famous spin -- flavour symmetries of HQET
(Isgur \& Wise 1989). They lead to relations among different
heavy-hadron form factors.
\item
From the Lagrangian $\bar h_v i v\cdot D h_v$ the Feynman rules
for HQET can be read off. The propagator is 
\begin{equation}\label{hqetprop}
\frac{i}{v\cdot k}\frac{1+\not v}{2}
\end{equation}
where the projector $(1+\not\! v)/2$ appears since $h_v$ is a
constrained spinor (see (\ref{hvdef})). The interaction of the
heavy-quark field $h_v$ with gluons is given by the vertex
\begin{equation}\label{hqetvert}
i g v^\mu T^a
\end{equation}
These Feynman rules enter in the computation of QCD quantum corrections.
\item
The remaining terms in (\ref{lhqet1}) are the leading power corrections.
They have an intuitive interpretation. In the first term one
recognizes the operator for the nonrelativistic kinetic energy
$\vec p^2/(2m)$, which describes the residual motion of the heavy quark
recoiling against the light degrees of freedom inside the heavy hadron.
The last term represents the chromomagnetic interaction of the
heavy-quark spin with the gluon cloud. Both effects violate flavour
symmetry, the chromomagnetic term also spin symmetry, but they
are power suppressed. 
\item
So far we have only considered QCD interactions. Weak interactions
introduce external currents, which can also be incorporated in HQET.
A generic heavy-light transition current $\bar q\Gamma\Psi$,
arising for instance in semileptonic decays, can be represented as
\begin{equation}\label{qpsiqhv}
\bar q\,\Gamma\,\Psi=\bar q\,\Gamma\, h_v+{\cal O}(\frac{1}{m})
\end{equation}
replacing the heavy-quark field $\Psi$ by the HQET field $h_v$
using (\ref{psihv}).
\end{itemize}

\subsection{Theory of heavy-hadron masses}

Before considering HQET in the context of weak decays, let us
discuss a first application of the basic HQET Lagrangian
(\ref{lhqet1}) in the spectroscopy of heavy hadrons.
To be specific, we shall analyze the masses of the ground-state
mesons $B$ and $B^*$. These mesons constitute a doublet that arises because
the spin $1/2$ of the heavy quark couples with the {\it total spin\/}
$1/2$ of the light degrees of freedom in their ground state to
form a spin-$0$ and a spin-$1$ meson, the pseudoscalar $B$ and
the vector $B^*$, respectively. Because the $b$-quark spin decouples
in the heavy-quark limit, the state of the light cloud is identical
for $B$ and $B^*$ to leading order, and the angular-momentum coupling
described above is the appropriate scheme.
If we neglect the power corrections in (\ref{lhqet1}), we can 
immediately write down the composition of the meson masses
\begin{equation}\label{mbb0}
m^{(0)}_B=m^{(0)}_{B^*}=m_b+\bar\Lambda
\end{equation}
Evidently the meson mass has a component $m_b$ from the heavy
quark. In addition it has a term $\bar\Lambda={\cal O}(\Lambda_{QCD})$
from the energy of the light constituents. The latter is determined
only by the interactions among the light degrees of freedom
and their interaction with the static $b$-quark ($h_v$) through the
first term in (\ref{lhqet1}). It is therefore independent of $m_b$.
The sum of $m_b$ and $\bar\Lambda$ is a physical quantity, however,
separately both parameters are dependent on the scheme used to define them.

In order to include the first power corrections, we treat the
$1/m$ terms in (\ref{lhqet1}) as perturbations to the lowest-order
HQET dynamics. To first order in perturbation theory the corrections
to (\ref{mbb0}) are then simply given by the expectation values of the
$1/m$ terms. The proper normalization is obtained as follows.
If ${\cal H}=-{\cal L}_{1/m}$ is the Hamiltonian (density) corresponding
to the correction term ${\cal L}_{1/m}$ in (\ref{lhqet1}), and 
$H=\int d^3x\, {\cal H}$ is the Hamilton operator, the mass correction
due to ${\cal H}$ is just
\begin{equation}\label{dmbb1} 
\delta m_B=\langle B_1|H|B_1\rangle
\end{equation}
where $|B_1\rangle$ is the $B$-meson state normalized to one,
$\langle B_1|B_1\rangle=1$.
Using the conventionally normalized states with 
$\langle B|B\rangle=2 m_B V$, we can write
\begin{equation}\label{dmbhb} 
\delta m_B=
\frac{1}{2 m_B V}\langle B|\int d^3x\, {\cal H}(\vec x)|B\rangle=
\frac{1}{2 m_B V}\int d^3x\,\langle B|{\cal H}(0)|B\rangle=
\frac{\langle B|{\cal H}(0)|B\rangle}{2 m_B}
\end{equation}
where we have used the translation invariance of ${\cal H}$
and $\int d^3x = V$.
Defining
\begin{equation}\label{l1l2}
\lambda_1\equiv\frac{\langle B|\bar h(iD)^2h|B\rangle}{2 m_B}
\qquad\quad
\lambda_2\equiv\frac{1}{6}
\frac{\langle B|\bar h g\sigma\cdot Gh|B\rangle}{2 m_B}
\end{equation}
we obtain
\begin{equation}\label{dmbl1l2}
\delta m_B=-\frac{\lambda_1+3 \lambda_2}{2 m_b}
\end{equation}
Note that we may replace $D^2$ by $D^2_\perp$ in the definition of
$\lambda_1$, up to higher order corrections 
(see (\ref{dperp}), (\ref{heom1})).
The parameter $\lambda_1$ corresponds to (minus) the expectation value
of the momentum squared of the heavy quark,
$\lambda_1=- \langle \vec p^2_h\rangle={\cal O}(\Lambda^2_{QCD})$.
This gives a positive correction in (\ref{dmbl1l2}) representing
the (small) kinetic energy of the heavy-quark. The 
$\lambda_2$-correction to the mass reflects the interaction energy
of the heavy-quark spin with its hadronic environment, as already
discussed in the previous section.
While the $\lambda_1$-term is independent of the heavy-quark spin and
identical for $B$ and $B^*$, the chromomagnetic correction
$\sim\lambda_2={\cal O}(\Lambda^2_{QCD})$ is different for $B^*$.
We have
\begin{equation}\label{dmbsl1l2}
\delta m_{B^*}=-\frac{\lambda_1- \lambda_2}{2 m_b}
\end{equation}
Including (\ref{mbb0}) we arrive at the following expansion
for the meson masses
\begin{eqnarray}
m_B &=& m_b+\bar\Lambda - \frac{\lambda_1+3 \lambda_2}{2 m_b}\label{mbhq}\\
m_{B^*} &=& m_b+\bar\Lambda - \frac{\lambda_1- \lambda_2}{2 m_b}\label{mbshq}
\end{eqnarray}
where the dependence on $m_b$ is explicit order by order.

If we apply the heavy-quark limit to $D$ mesons, we obtain
analogous relations
\begin{eqnarray}
m_D &=& m_c+\bar\Lambda - \frac{\lambda_1+3 \lambda_2}{2 m_c}\label{mdhq}\\
m_{D^*} &=& m_c+\bar\Lambda - \frac{\lambda_1- \lambda_2}{2 m_c}\label{mdshq}
\end{eqnarray}
with the same $\bar\Lambda$, $\lambda_1$ and $\lambda_2$ as before.

These results have a few interesting consequences.
First, $\lambda_2$ parametrizes the spin-splitting between the
pseudoscalar and the vector mesons:
\begin{eqnarray}
m_{B^*}-m_B &=& \frac{2\lambda_2}{m_b}=46\,{\rm MeV}\label{mbsmb}\\
m_{D^*}-m_D &=& \frac{2\lambda_2}{m_c}=141\,{\rm MeV}\label{mdsmd}
\end{eqnarray}
HQET predicts that the spin-splitting scales inversely proportional
to the heavy-quark mass. This is seen to be quite well fulfilled
given that $m_b\approx 3 m_c$.
Relation (\ref{mbsmb}) can be used to determine the nonperturbative
quantity $\lambda_2$ from experiment
\begin{equation}\label{l2num}
\lambda_2=\frac{1}{4}(m^2_{B^*} - m^2_B)=0.12\,{\rm GeV}^2
\end{equation}
On the other hand, the quantity $\lambda_1$ has to be estimated
theoretically.
Finally one may introduce the spin-averaged masses
\begin{eqnarray}
\bar m_B &\equiv& \frac{m_B+3 m_{B^*}}{4}=
m_b+\bar\Lambda-\frac{\lambda_1}{2 m_b} \label{mbavg}\\
\bar m_D &\equiv& \frac{m_D+3 m_{D^*}}{4}=
m_c+\bar\Lambda-\frac{\lambda_1}{2 m_c} \label{mdavg}
\end{eqnarray}
This eliminates $\lambda_2$ and yields the useful result
\begin{equation}\label{mbmcmbmd}
m_b-m_c=(\bar m_B - \bar m_D)
\left( 1 -\frac{\lambda_1}{2 \bar m_B \bar m_D}\right)
\end{equation}
Since the $\lambda_1$-correction is fairly small, the quark-mass
difference is rather well determined, much better than individual
quark masses.

\subsection*{Exercise}

{\it Derive the relative factor between the chromomagnetic
correction to the mass of the $B$ and the $B^*$ meson.}

Solution: Denote the heavy-quark spin by $\vec s$, the
total spin of the light degrees of freedom by $\vec j$ and the
total spin of the meson by $\vec J=\vec s+\vec j$. The chromomagnetic
field of the light cloud has to be proportional to $\vec j$.
Hence the energy of the interaction between this field and $\vec s$
is proportional to $\langle \vec s\cdot \vec j\rangle$.
Angular momentum algebra gives 
$\langle 2\vec s\cdot \vec j\rangle =J(J+1)-s(s+1)-j(j+1)$,
which is $(-3/2)$ for $B$ and $(1/2)$ for $B^*$, hence
the relative factor $(-1/3)$ of the $\lambda_2$-term
in (\ref{dmbsl1l2}) with respect to (\ref{dmbl1l2}).

\subsection{Heavy-light currents and $f_B$}

The $B$-meson decay constant $f_B$ is defined by the matrix element
\begin{equation}\label{fbdef}
\langle 0|A_\mu|B(p)\rangle =-i f_B m_B v_\mu
\end{equation}
of the heavy-light axial vector current
\begin{equation}\label{amu5}
A_\mu\equiv\bar q\gamma_\mu\gamma_5\Psi
\end{equation}
Here $q$ is the light-quark, $\Psi$ the heavy-quark field
in full QCD, with $\Psi=b$ in the present case. The $B$-meson
momentum is $p=m_B v$.

Let us analyze $A_\mu$ in HQET, including QCD corrections.
The expansion of $A_\mu$ in HQET to leading order in $1/m$,
but allowing for QCD effects, has the form
\begin{equation}\label{aa1a2}
A=C_1(\mu) \tilde A_1+C_2(\mu) \tilde A_2+{\cal O}(\frac{1}{m})
\end{equation}
\begin{equation}\label{a1a2t}
\tilde A_1=\bar q\gamma_\mu\gamma_5 h_v\qquad \quad
\tilde A_2=\bar q v_\mu\gamma_5 h_v
\end{equation}
The matching conditions at the $b$-quark mass scale $\mu=m_b$ are
\begin{equation}\label{c1c2mb}
C_1(m_b)=1+{\cal O}(\alpha_s) \qquad C_2(m_b)={\cal O}(\alpha_s)
\end{equation}
To leading order in QCD only $\tilde A_1$ is present in HQET,
with coefficient one. Radiative corrections at ${\cal O}(\alpha_s)$
modify $C_1$ and generate a new operator $\tilde A_2$.
Note that the matching calculation of the full-QCD current $A$ onto
HQET, leading to (\ref{aa1a2}), is completely analogous to the OPE
procedure of constructing the effective weak Hamiltonian from the
$W$-exchange amplitude in the full standard model, which we have
discussed in sec. 3. The difference is only that a $1/M_W$ expansion
is performed in the latter case, and a $1/m_b$ expansion in the case of 
HQET. The basic philosophy is essentially the same. In particular,
a factorization of long and short-distance contributions is obtained:
Contributions from large scales $>\mu$, including the $m_b$-dependence,
is again contained in the coefficient functions $C_{1,2}$. Soft scales
$<\mu$ are factorized into the hadronic matrix elements of
$\tilde A_{1,2}$.

In contrast to the full-QCD current $A$, the HQET currents do have
an anomalous dimension, reflecting a logarithmic dependence
of $f_B$ on the heavy-quark mass at ${\cal O}(\alpha_s)$.
The logarithms can be resummed by renormalization group methods,
again in full analogy to the procedure in sec. 3. In leading
logarithmic approximation (LLA) $C_2$ can be neglected and $C_1$
acquires the familiar form
\begin{equation}\label{c1mua}
C_1(\mu)=\left[\frac{\alpha_s(m_b)}{\alpha_s(\mu)}\right]^{-2/\beta_0}
\end{equation}
Here the LLA assumes the hierarchy $\alpha_s(m_b)\ll 1$,
$\alpha_s\ln(m_b/\mu)={\cal O}(1)$, which holds in the heavy-quark
limit ($m_b$ large, $\mu={\cal O}(1\,{\rm GeV})$).

To express $f_B$ in HQET via (\ref{fbdef}), (\ref{aa1a2}) and
(\ref{c1mua}), we need the matrix element of $\tilde A_1$
\begin{equation}\label{ftildef}
\langle 0|\tilde A_1|B(p)\rangle =-i \tilde f(\mu) \sqrt{m_B} v_\mu
\end{equation}
Since the dynamics of HQET is independent of $m_b$, the reduced
decay constant $\tilde f(\mu)$ is $m_b$-independent. The only
$m_b$-dependence in (\ref{ftildef}) enters through a trivial factor
$\sqrt{m_B}$ from the normalization of the $B$-meson state, which in the
usual convention is given by
\begin{equation}\label{bnorm}
\langle B|B\rangle=2 m_B V
\end{equation}
Collecting the ingredients, (\ref{aa1a2}) yields
\begin{equation}\label{fbftil}
f_B=\frac{\tilde f(\mu)}{\sqrt{m_B}}
\left[\frac{\alpha_s(m_b)}{\alpha_s(\mu)}\right]^{-2/\beta_0}
\end{equation}
This expression for $f_B$ is true to leading order in the HQET expansion
in $\Lambda_{QCD}/m_b$ and in leading logarithmic approximation in QCD.
$\tilde f(\mu)$ in (\ref{fbftil}) is still a nonperturbative quantity
to be determined by other methods. However, the dependence of $f_B$
on the heavy-quark mass is now explicit.
Eq. (\ref{fbftil}) implies the scaling behaviour $f_B\sim 1/\sqrt{m_B}$,
up to a calculable logarithmic dependence on $m_b$. In principle such 
a relation can be used to relate $f_B$ to the analogous quantity
$f_D$ for heavy mesons with charm. In practice, it turns out that the
leading order scaling result for $f_B$ is not very well fulfilled
even for the $b$-mass scale and that subleading power corrections are 
important in this case. Nevertheless the result in (\ref{fbftil})
is of conceptual interest and can serve as a simple example of an
application of HQET.

\subsection{Heavy-heavy currents: $\bar B\to D^{(*)}l\bar\nu$ and $V_{cb}$}

One of the most important results of HQET is the extraction of $V_{cb}$
from exclusive semileptonic $\bar B\to D^*l\bar\nu$ decay.
We will here give a short outline of the main steps in this analysis.
Starting point is the differential decay rate
\begin{equation}\label{dgbdln}
\frac{d\Gamma(\bar B\to D^*l\bar\nu)}{dw}=|V_{cb}|^2
\, {\cal K}(w)\, {\cal F}^2(w)
\end{equation}
in the kinematical variable $w=v\cdot v'$, where $v$ and $v'$ are the
4-velocities of $\bar B$ and $D^*$, respectively. The dependence
of (\ref{dgbdln}) on $|V_{cb}|$, the quantity of interest, is obvious, 
and ${\cal K}(w)$ is a known kinematical function.
Finally, ${\cal F}(w)$ contains the nontrivial QCD dynamics encoded
in the $\bar B\to D^*$ transition form factors. The corresponding
matrix elements of the weak currents can be written in the heavy-quark
limit as
\begin{eqnarray}
\frac{1}{\sqrt{m_{D^*} m_B}}
\langle D^*(v',\epsilon)|\bar c\gamma_\mu b|\bar B(v)\rangle &=&
i\, \xi(w)\, \varepsilon(\mu,\epsilon,v',v) \\
\frac{1}{\sqrt{m_{D^*} m_B}}
\langle D^*(v',\epsilon)|\bar c\gamma_\mu\gamma_5 b|\bar B(v)\rangle &=&
\xi(w)\, \left[(1+w)\epsilon_\mu-(\epsilon\cdot v)v'_\mu\right]
\end{eqnarray}
In the heavy-quark limit, that is to lowest order in HQET, all
hadronic dynamics is expressed in a single function $\xi(w)$,
the Isgur--Wise function (Isgur \& Wise 1989).
In this limit we further have
\begin{equation}\label{fwxiw}
{\cal F}(w)=\xi(w)
\end{equation}
Moreover, $\xi$ is absolutely normalized at the no-recoil point
\begin{equation}\label{xi1}
\xi(1)=1
\end{equation}
The no-recoil point $w=1$ corresponds to the kinematical situation
where the $D^*$ meson stays at rest in the rest frame of the decaying
$\bar B$ ($v'=v\, \Rightarrow\, w=1$). Measuring $d\Gamma/dw$ at $w=1$,
$|V_{cb}|$ can then be determined from (\ref{dgbdln}) since all
ingredients are known.
Because $w=1$ is at the edge of phase space, an extrapolation is
necessary to find $d\Gamma/dw|_{w=1}$ from the measured spectrum.

For a realistic analysis corrections to the heavy-quark limit need
to be considered. An important property of $\bar B\to D^* l\bar\nu$
is that linear power corrections in HQET are absent, $\delta_{1/m}=0$,
where $m$ can be either $m_c$ or $m_b$. Consequently the leading
corrections enter only at second order and are thus greatly reduced.
This result is known as Luke's theorem. The absence of linear 
corrections does not hold for $\bar B\to D l\bar\nu$ decays, hence
the particular importance of $\bar B\to D^* l\bar\nu$.
Including corrections, the lowest order approximation
${\cal F}(1)=\xi(1)=1$ is modified to
\begin{equation}\label{f1etaa}
{\cal F}(1)=\eta_A(1+\delta_{1/m^2})
\end{equation}
where $\delta_{1/m^2}$ are the second order power corrections
and $\eta_A$ is a correction from perturbative QCD. To first order
in $\alpha_s$ it reads
\begin{equation}\label{etaalphas}
\eta_A=1+\frac{\alpha_s}{\pi}
\left(\frac{m_b+m_c}{m_b-m_c}\ln\frac{m_b}{m_c}-\frac{8}{3}\right)
\end{equation}
A detailed numerical analysis yields (Harrison \& Quinn 1998)
\begin{equation}\label{f1num}
{\cal F}(1)=0.913\pm 0.042
\end{equation}
which gives (H\"ocker et al. 2001)
\begin{equation}\label{vcbexcl}
V_{cb}=0.0409\pm 0.0014_{exp} \pm 0.0019_{th}
\end{equation}

To summarize the crucial points for the extraction of $V_{cb}$
from $\bar B\to D^* l\bar\nu$ decay:

\begin{itemize}
\item
Heavy-quark symmetry relates the various semileptonic form factors
(four different functions $V$, $A_0$, $A_1$, $A_2$ in full QCD)
to a single quantity $\xi(w)$, the Isgur--Wise function.
\item
The function $\xi$ is absolutely normalized, $\xi(1)=1$.
This property has an intuitive reason: At the kinematical point
$w=1$ the decaying $b$-quark at rest is transformed into a $c$-quark,
also at rest. Since both quarks are treated in the static approximation
($m_b$, $m_c\to\infty$, $m_b/m_c$ fixed), the light hadronic cloud
doesn't notice the flavour change $b\to c$ and is transfered from the
$\bar B$ to a $D$ meson with probability one. The function $\xi$
is identical for $\bar B\to D$ and $\bar B\to D^*$ transitions, because
these are related by heavy-quark spin symmetry.
\item
HQET provides a framework for systematic corrections to the strict
heavy-quark limit governed by $\xi(w)$. Luke's theorem guarantees the
absence of first-order corrections in $1/m$ for $\bar B\to D^* l\bar\nu$.
\end{itemize}

\subsection{HQET -- conclusions}

We would finally like to summarize the basic ideas and virtues of
HQET, and to re-emphasize the salient points.

\begin{itemize}
\item
HQET describes the static approximation for a heavy quark,
covariantly formulated as an effective field theory and allowing
for a systematic inclusion of power corrections.
\item
Order by order in the expansion in $\Lambda_{QCD}/m$ HQET
achieves a {\it factorization\/} of hard, perturbative contributions
(momentum scales between $m$ and a factorization scale $\mu$) and
soft, nonperturbative contributions (scales below $\mu$).
The former are contained in Wilson coefficients, the latter in the matrix
elements of HQET operators.
\item
The procedure of matching full QCD onto HQET is analogous to the
construction of the effective weak Hamiltonian ${\cal H}_{eff}$.
The difference lies in the massive degrees of freedom that are being
integrated out: the $W$ boson (mass $M_W$) for ${\cal H}_{eff}$,
the lower-component spinor field $H_v$ (mass $2m$) for HQET.
The perturbative matching can be supplemented by RG resummation of
logarithms, $\ln(M_W/\mu)$ in the former case, $\ln(m/\mu)$ in the 
latter.
\item
The usefulness of HQET is based on two important features:
The spin-flavour symmetry of HQET relates form factors in the 
heavy-quark limit and thus reduces the number of unknown hadronic
quantities. The dependence on the heavy-quark masses is made
explicit (scaling, power corrections).
\end{itemize}

We conclude with briefly mentioning another field, called
large energy effective theory (LEET), which has some similarities
with HQET.
LEET is needed for matrix elements of the form
$\langle M|\bar q\,\Gamma\, b|\bar B\rangle$ at 
{\it large recoil\/} of the light meson $M=\pi$, $\rho$, $K^{(*)}$,
$\ldots$. HQET is not sufficient in this situation because not only
soft but also collinear infrared singularities need to be factorized.
The latter occur due to the light-like kinematics of the fast and
energetic light quark emitted from the weak current.
To define LEET the usual heavy-quark limit can be considered for the
$B$ meson with velocity $v$. The large-energy limit is taken for the
light meson $M$ with light-like momentum vector $E n$.
Here $E={\cal O}(m_b)$ is the energy of $M$ and $n$ is a light-like
4-vector with $n^2=0$ and $v\cdot n=1$. The momentum of the
energetic light quark $q$ is written as $p_q=En+k$, with a residual
momentum $k={\cal O}(\Lambda_{QCD})$. In formal analogy to the fields
$h_v$ and $H_v$ in HQET, the new light-quark fields
\begin{equation}\label{qnq}
q_n(x)=e^{i E n\cdot x}\,\frac{\not n\not v}{2}q(x)
\qquad Q_n(x)=e^{i E n\cdot x}\,\frac{\not v\not n}{2}q(x)
\end{equation}
can be defined and used in the construction of LEET
(Dugan \& Grinstein 1991, Charles et al. 1999, Beneke \& Feldmann 2001,
Bauer et al. 2001a).
As a consequence of the LEET limit the ten form factors needed to
describe all matrix elements
$\langle M|\bar q\,\Gamma\, b|\bar B\rangle$
of bilinear heavy-light currents can be reduced to only three 
independent functions. LEET has received increasing interest
quite recently and is still under active development.

\section{Inclusive decays and the heavy quark expansion}

\subsection{Basic formalism and theory of lifetimes}

The heavy-quark limit, $m\gg\Lambda_{QCD}$, proves to be extremely
useful also for the computation of {\it inclusive\/} decay rates
of heavy hadrons (Chay et al. 1990, Bigi et al. 1992, 1997). 
The specific technique appropriate for this
application is distinct from HQET and goes by the name of
heavy quark expansion (HQE).
Consider the total decay rate $\Gamma_H$ of a heavy hadron $H$.
Starting point for the HQE is the following representation of
$\Gamma_H$
\begin{equation}\label{ght}
\Gamma_H=\frac{1}{2 m_H}\langle H|{\cal T}|H\rangle\equiv
\langle{\cal T}\rangle
\end{equation}
where the transition operator ${\cal T}$ is defined as
\begin{equation}\label{timhh}
{\cal T}=
{\rm Im}\,\, i\int d^4x\, T\, {\cal H}_{eff}(x){\cal H}_{eff}(0)
\end{equation}
with ${\cal H}_{eff}$ the effective weak Hamiltonian.
Eqs. (\ref{ght}), (\ref{timhh}) express the total decay rate as
the absorptive part of the forward scattering amplitude $H\to H$
under the action of ${\cal H}_{eff}$. This expression is refered to
as the optical theorem by analogy to a similar relation in optics.
One may rewrite (\ref{ght}), (\ref{timhh}) in a more directly
understandable form by inserting a complete set of states
$|X\rangle\langle X|$ between the two factors of ${\cal H}_{eff}$
in (\ref{timhh}) and removing the $T$-product by explicitly taking the
absorptive part. This yields
\begin{equation}\label{ghhxxh}
\Gamma_H\sim\langle H|{\cal H}_{eff}|X\rangle
     \langle X|{\cal H}_{eff}|H\rangle
\end{equation}
where one immediately recognizes the decay rate as the modulus
squared of the decay amplitude (summed over all final states $X$).
The reason to introduce (\ref{timhh}) is that the $T$-product, by
means of Wick's theorem, allows for a direct evaluation in terms
of Feynman diagrams.

\begin{figure}
[htbp]
\centering
\includegraphics[width=15cm,height=7cm]{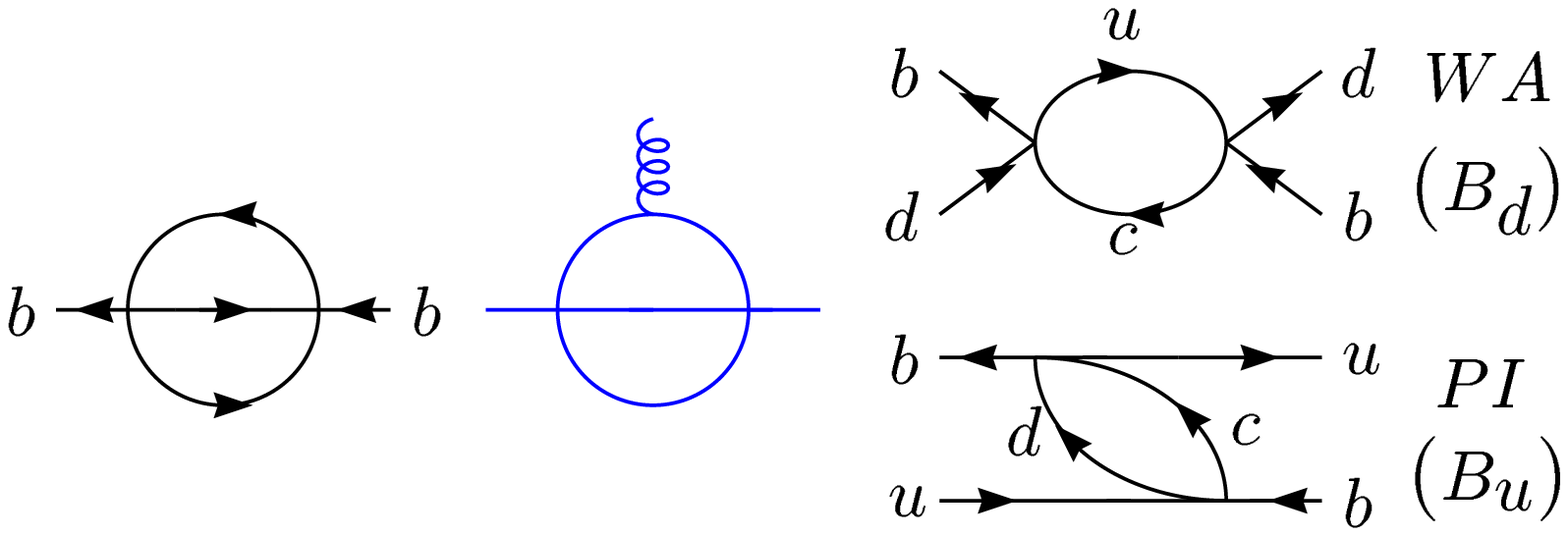}
\vspace*{-1cm}
\caption{Heavy quark expansion for the total decay rate
of $b$-hadrons.}
\label{fig:thqe}
\end{figure}

In order to compute $\Gamma_H$ an operator product expansion
is applied to (\ref{timhh}), resulting in a series of local
operators of increasing dimension. The coefficients of these
operators are correspondingly suppressed by increasing powers of
$1/m_b$. The series has the form
\begin{equation}\label{thqe}
{\cal T}=\Gamma_b\, \bar bb
+\frac{z_G}{m^2_b}\,\bar b g\sigma\cdot Gb
+\sum\frac{z_{qi}}{m^3_b}\,\bar b\Gamma_i q\,\,\bar q\,\Gamma_i b
+\ldots
\end{equation}
where we have written the first few operators of dimension
three ($\bar bb$), five ($\bar b g\sigma\cdot Gb$) and
six ($\bar b\Gamma_i q\,\,\bar q\,\Gamma_i b$). The matrix
elements of the operators contain the soft, nonperturbative physics,
their Wilson coefficients $\Gamma_b$, $z_k$ the hard contributions,
which are calculable in perturbation theory.
Again, the coefficients are determined by an appropriate matching
calculation between (\ref{timhh}) and the r.h.s. of (\ref{thqe}).
The Feynman diagrams for the three terms in (\ref{thqe}) are shown
in Fig. \ref{fig:thqe}
The two weak-interaction vertices in these diagrams correspond
to the two factors of ${\cal H}_{eff}$ in the definition of
${\cal T}$ in (\ref{timhh}) (the absorptive part of the diagrams
is understood).

Obviously, the heavy quark expansion is different from HQET.
However, we may still use HQET in conjunction with (\ref{thqe})
in order to further analyse the hadronic matrix elements.
An important example is the leading dimension-three operator
$\bar bb$. Its matrix element between heavy-hadron states $H$
can be expanded in HQET as 
\begin{equation}\label{bbarb}
\langle\bar bb\rangle=1+\frac{1}{2m^2_b}\,\langle\bar h(iD)^2h\rangle
+\frac{1}{4m^2_b}\,\langle\bar h g\sigma G h\rangle
\end{equation}
where $\langle\ldots\rangle\equiv\langle H|\ldots |H\rangle/(2 m_H)$.
\begin{itemize}
\item
Eqs. (\ref{ght}), (\ref{thqe}) and (\ref{bbarb}) imply that to leading
order in the HQE $\Gamma_H =\Gamma_b$, that is the total decay
rate of all $b$-flavoured hadrons is equal to the rate of free
$b$-quark decay. Pictorially this can be seen from the first diagram
in Fig. \ref{fig:thqe}, which represents essentially the amplitude
squared for the partonic decay of a $b$-quark.
Note also that perturbative QCD corrections to $\Gamma_b$ can
consistently be taken into account. The gluonic corrections to inclusive
$b$-quark decay are infrared safe, as required for $\Gamma_b$
in its role as a Wilson coefficient of the HQE. Also, corrections
proportional to powers of $\alpha_s(m_b)\sim 1/\ln(m_b/\Lambda)$
are only suppressed by inverse powers of $\ln m_b$ in the heavy-quark
limit, and hence formally leading in comparison to higher corrections
in the HQE, which are suppressed by powers of $\Lambda/m_b$.
The calculation of heavy-quark decay in the parton picture has been
used since the beginnings of heavy-quark physics as an approximation
for inclusive decays of the corresponding heavy hadrons. As we have
seen, the HQE gives a formal justification for this approach and
provides us with a theoretical framework to compute nonperturbative 
corrections.
\item
The first correction term in (\ref{bbarb}) depends on the expectation 
value of the momentum squared $\langle\vec p^2\rangle$ of the
heavy quark inside the hadron. This matrix element is non-zero
because the heavy quark is recoiling against the light degrees of freedom
through gluonic interactions in the hadronic bound state. This term
has a very intuitive interpretation. It corresponds to a correction
factor $1-\langle\vec p^2\rangle/(2 m^2_b)$ $=$ 
$1-\langle\vec v^2_b\rangle/2$, which is just the reduction of the free
decay rate from time dilatation due to the recoil motion of the
heavy quark. The second correction comes from interactions of the
light hadronic cloud with the heavy-quark spin. We have
\begin{equation}\label{hsigmagh}
\langle\bar h g\sigma G h\rangle =
\left\{\begin{array}{cl}
        \frac{3}{2}(m^2_{B^*}-m^2_B) & H=B\\
                    0                & H=\Lambda_b
       \end{array}\right.
\end{equation}
The result is zero for the $\Lambda_b$ baryon since the light
degrees of freedom are in a state of zero total angular momentum.
Note that the spin interaction enters twice in (\ref{thqe}), explicitly
with coefficient $z_G$ and via the expansion of $\langle\bar bb\rangle$.
\item
The leading nonperturbative corrections start only at second order.
There is no correction linear in $1/m_b$. This is because there
is no gauge-invariant operator of dimension four that could
appear in the HQE.
\item
At order $1/m^3_b$ contributions appear where the spectator quark
participates directly in the weak interactions.
For $b$-mesons they can be interpreted as the effect of weak
annihilation of the $b$-quark with the valence $\bar d$-quark
(for $\bar B_d$) and as the effect of Pauli interference (for $\bar B_u$).
The latter phenomenon occurs because in the nonleptonic decay of
a $\bar B_u$, $b(\bar u)\to c\bar ud(\bar u)$, two identical
$\bar u$-quarks are present in the final state. These
corrections distinguish in particular among $B_d$ and $B_u$ mesons
and are responsible for their lifetime difference.
Despite the suppression by three powers of $m_b$ these effects can be
relatively important due to their two-body kinematics, which brings
a phase-space enhancement factor of $16\pi^2$ in comparison to the
leading three-body decay. 
\end{itemize}

As one of the possible applications, the HQE provides us with a theory
of heavy-hadron lifetimes. The deviations of lifetime ratios from
unity probes the power corrections. At present there are still
sizeable theoretical uncertainties due to the hadronic matrix 
elements $\langle\bar b\,\Gamma q\, \bar q\,\Gamma b\rangle$.
They can in principle be computed with the help of lattice gauge
theory. Table 2 shows a comparison of theoretical predictions and 
experimental results (see for instance (Ligeti 2001)).
\begin{table}[h]
\label{tab:lifetimes}
\caption[]{}
\begin{center}
\vspace*{0.4cm}
\begin{tabular}{|c|c|c|}
\hline
 & {exp.} & th.\\
\hline
$\tau(B^+)/\tau(B^0_d)$ & $1.068\pm 0.016$ & $1 - 1.1$\\
\hline
$\bar\tau(B_s)/\tau(B_d)$ & $0.947\pm 0.038$ & $0.99 - 1.01$\\
\hline
$\tau(\Lambda_b)/\tau(B_d)$ & $0.795\pm 0.053$ & $0.9 - 1.0$\\
\hline
\end{tabular}
\end{center}
\end{table}

\subsection{Local quark-hadron duality}

A systematic uncertainty within the HQE framework, which is
often debated in the literature, arises from the issue of
quark-hadron duality. In this paragraph we give a brief and
heuristic discussion of the basic idea behind this topic.

\begin{figure}
[htbp]
\centering
\vspace*{-2.5cm}
\includegraphics[width=10cm,height=16cm]{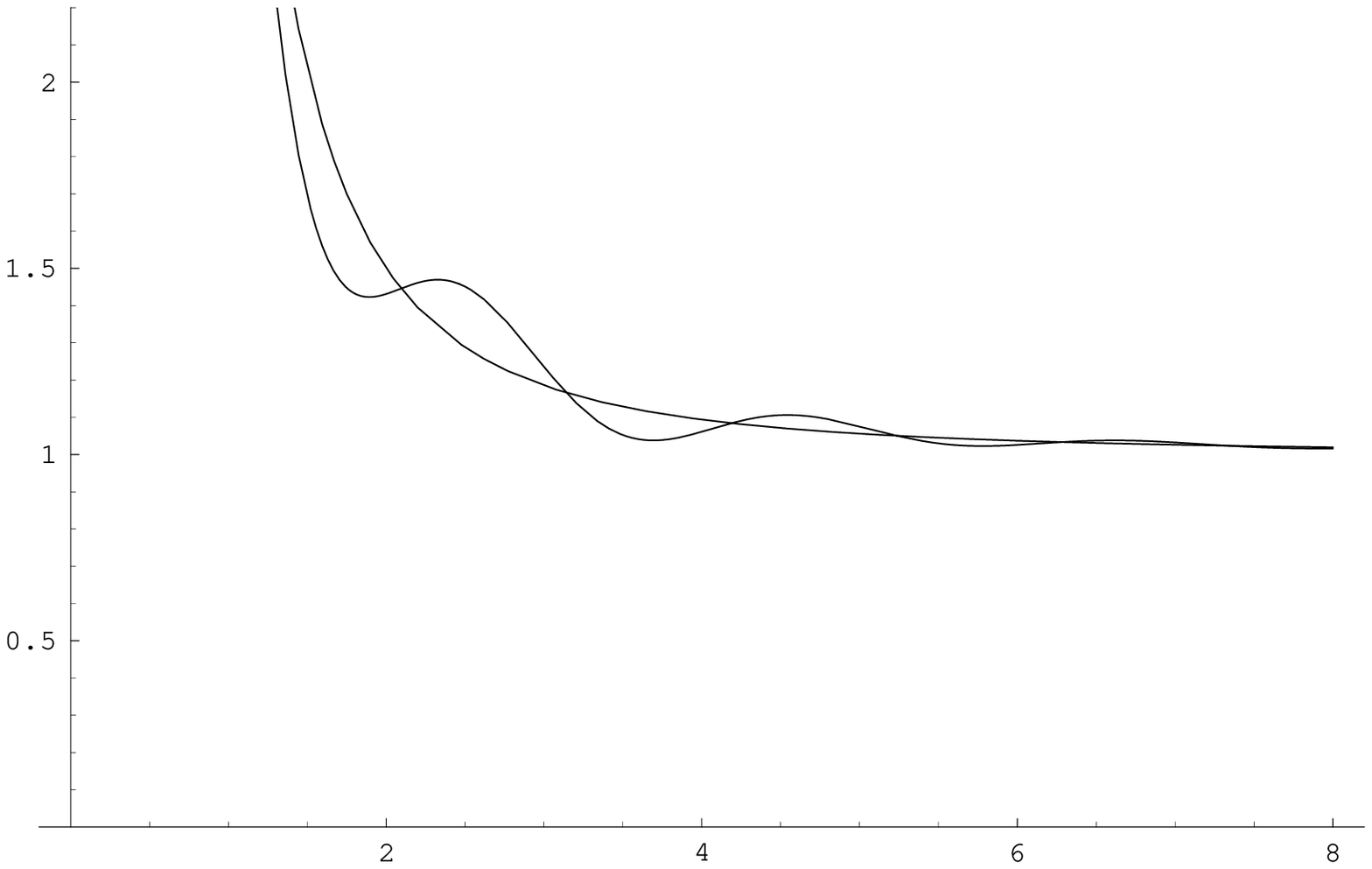}
\vspace*{-3.5cm}
\caption{$\Gamma/\Gamma_0$ as function of $m_b$.}
\label{fig:lqhd}
\end{figure}

The theoretical prediction for an inclusive decay rate obtained
from the HQE has the form
\begin{equation}\label{hqeseries}
\Gamma/\Gamma_0=1+\sum^\infty_{n=2}z_n\,\left(\frac{\Lambda}{m_b}\right)^n
\end{equation}
where we have denoted the leading, free-quark result by $\Gamma_0$.
Let us consider the decay rate as a function of $m_b$, keeping
$\Lambda=\Lambda_{QCD}$ constant.
Then the quantity $\Gamma/\Gamma_0$, to any finite order in 
$(\Lambda/m)$, is a simple  polynomial expression in this variable.
This is sketched as the monotonous curve in Fig. \ref{fig:lqhd}
showing $\Gamma/\Gamma_0$ as function of $m_b$ (in arbitrary units).
Now, since by construction the HQE for $\Gamma/\Gamma_0$ yields
a power expansion in $(\Lambda/m)$, any term of the form
\begin{equation}\label{expsin}
\exp\left(-\left(\frac{m_b}{\Lambda}\right)^k\right)
\sin\left(\frac{m_b}{\Lambda}\right)^k
\end{equation}
for example, present in the true result for $\Gamma/\Gamma_0$
would be missed by the HQE. This is due to the exponential suppression
in the expansion parameter. In fact, the function $\exp(-1/x)$ is
non-analytic. Its power expansion around $x=0$ gives identically
zero. However, such (or similar) terms are expected to be part of the true
$\Gamma/\Gamma_0$ on general grounds.
The corresponding complete result for $\Gamma/\Gamma_0$,
including such a term, is sketched as the oscillating graph
in Fig. \ref{fig:lqhd}.
This true curve represents the physical result for the decay rate
$\Gamma/\Gamma_0$, which consists of the inclusive sum over all
the different exclusive decay channels. It is intuitively understandable 
that the true $m_b$-dependence will have such a damped oscillating 
behaviour:
If we imagine to continually increase $m_b$, $\Gamma/\Gamma_0$
will undergo a small jump whenever it reaches a value at which the
presence of a further higher hadronic resonances in the final state
becomes kinematically allowed. Since the excited hadrons have finite
widths, the threshold behaviour will be smoothed out, resulting in the
pattern of damped oscillations.

The term {\it quark-hadron duality\/} refers to the idea that the
inclusive rate as the sum over all exclusive {\it hadronic\/} decay channels
and the inclusive rate as predicted by the heavy {\it quark\/} expansion
are dual to each other. This means they are both valid representations
of the same quantity using different descriptions, the hadron level
or the quark level. The term {\it local} refers to the fact that
the energy scale $m_b$ is a fixed quantity, as opposed to e.g. the
centre-of-mass energy in $e^+e^-$ annihilation, which can be averaged
to obtain suitably defined ``global'' quantities. 
In principle, the hadronic description gives
the true result, measured in experiment. The problem is, however,
that we would have to compute all exclusive rates first, which is
far beyond our current control of nonperturbative QCD.
On the other hand, the HQE calculation can be performed, within some
uncertainties, but it is clear that the result need not be identical
to the true answer. A deviation between the latter and the HQE
(including power corrections) 
is refered to as a violation of quark-hadron duality.  
Indeed, contributions violating quark-hadron duality are expected
(see(\ref{expsin})), but the numerical size of these terms cannot be 
strictly computed at present. Conceptually this is no problem
because they are formally subleading in comparsion to 
power corrections, so that the HQE still makes sense even at higher
orders. The remaining question is how large can violations of
quark-hadron duality be numerically. While there are at the moment,
within the uncertainties intrinsic to HQE,
no established cases in inclusive $B$ decays where duality 
is violated, the issue clearly needs further investigation,
both theoretically and phenomenologically.

A more detailed account of the status of quark-hadron duality 
can be found in the papers by Blok, Shifman \& Zhang (1998),
Shifman (2000) and Bigi \& Uraltsev (2001).

\subsection{Inclusive semileptonic decays -- $V_{ub}$, $V_{cb}$}

The HQE cannot only  be applied to the total decay rates,
but also to inclusive rates with specific flavour quantum
numbers in the final state, such as semileptonic processes.
Furthermore one can analyze differential decay rates.

An example of special interest is the inclusive decay
$\bar B\to X_c l\bar\nu$, which can be used to extract $V_{cb}$.
The HQE for the integrated rate has the form
\begin{equation}\label{bclnu}
\Gamma(B\to X_c l\nu)=\frac{G^2_F m^5_b}{192\pi^3}\, {|V_{cb}|^2}
\,\left[ z_3 \left(1+\frac{{\lambda_1}+3{\lambda_2}}{2 m^2_b}\right)
+ \, z_5\,\frac{6{\lambda_2}}{m^2_b}+\ldots\right]
\end{equation}
with
\begin{equation}\label{bbbgb}
\langle\bar bb\rangle=1+\frac{{\lambda_1}+3{\lambda_2}}{2 m^2_b}
\qquad\quad \langle\bar b\sigma G b\rangle=6{\lambda_2}=
\frac{3}{2}(m^2_{B^*}-m^2_B)
\end{equation}
The Wilson coefficients read
\begin{equation}\label{bslz3}
z_3=1-8x+8x^3-x^4-12 x^2 \ln x +{\cal O}(\alpha_s)
\end{equation}
\begin{equation}\label{bslz5}
z_5=-(1-x)^4
\end{equation}
Where $x=(m_c/m_b)^2$.

A major source of theoretical uncertainty for the determination
of $|V_{cb}|$ using (\ref{bclnu}) is the $b$-quark mass.
This appears to be especially problematic
since $m_b$ comes with the fifth power in (\ref{bclnu}).
Fortunately, however, the actual situation is not as bad. Taking 
into account the phase-space function $z_3$, one finds that the 
combined dependence on $m_b$ and $m_c$ shows the approximate behaviour
\begin{equation}\label{bslmbmc} 
\Gamma(B\to X_c l\nu)\sim m^{2.3}_b\, (m_b-m_c)^{2.7}
\end{equation}
Since the difference $m_b-m_c$ is better known than the
individual quark masses, the corresponding uncertainty is reduced.
The quark-mass difference is in fact constrained by HQET, which
gives (\ref{mbmcmbmd})
\begin{eqnarray}
m_b-m_c &=& (\bar m_B-\bar m_D)
\left(1-\frac{\lambda_1}{2\bar m_B\bar m_D}\right) \nonumber \\
 &=& 3.40\pm 0.03\pm 0.03 {\rm GeV}
\end{eqnarray}
where $\bar m_B\equiv (m_B+3 m_{B^*})/4$.

The QCD corrections to $z_3$ are known to 
${\cal O}(\alpha_s)$ and partly at ${\cal O}(\alpha^2_s)$.
The special class of corrctions ${\cal O}(\beta^{n-1}_0\alpha^n_s)$
has been calculated to all orders $n$.

Numerically the inclusive method gives (H\"ocker et al. 2001)
\begin{equation}\label{vcbincl}
V_{cb}=0.04076\pm 0.00050_{exp} \pm 0.00204_{th}
\end{equation}
which can be compared with the result from the
exclusive determination via $\bar B\to D^* l\bar\nu$.

One can also try to extract $|V_{ub}|$ from $B\to X_u l\nu$
decays. This is more difficult since the very large background
from semileptonic $b\to c$ transitions requires kinematical
cuts (in the lepton energy, the hadronic or the dilepton invariant
mass), which renders the HQE less reliable and introduces larger
uncertainties.
A recent discussion has been given by Bauer et al. (2001b).
The HQE has further useful applications, for instance in the case of
the inclusive rare decays
$B\to X_{s,d}\gamma$, $B\to X_{s,d}l^+l^-$, 
or $B\to X_{s,d}\nu\bar\nu$.

\subsection*{Exercise}

{\it Show that quark-hadron duality is {\rm exactly} fulfilled
for the semileptonic $b\to c$ transition rate in the
Shifman-Voloshin (small-velocity, or SV) limit 
$m_b$, $m_c$ $\gg$ $m_b-m_c$ $\gg\Lambda_{QCD}$.
This holds with only two exclusive channels on the hadronic
side of the duality relation, that is the inclusive rate
is saturated as
$\Gamma(B\to X_c l\nu)\equiv \Gamma(B\to D l\nu)+\Gamma(B\to D^* l\nu)$
in this limit.}

Solution:
We start from the exclusive differential decay rates
in the {\it heavy-quark limit}. 
They read (see e.g. Harrison \& Quinn 1998):
\begin{eqnarray}
\frac{d\Gamma(B\to D l\nu)}{dw}
&=& \frac{G^2_F |V_{cb}|^2}{48\pi^3}
(m_B+m_D)^2 m^3_D (w^2-1)^{3/2}\, \xi^2(w) \label{dgdwd}\\
\frac{d\Gamma(B\to D^* l\nu)}{dw}
&=& \frac{G^2_F |V_{cb}|^2}{48\pi^3}
(m_B-m_D^*)^2 m^3_{D^*} \sqrt{w^2-1}(w+1)^2 \nonumber \\
&& \cdot\left(
1+\frac{4w}{w+1}\frac{m^2_B-2 w m_B m_{D^*}+m^2_{D^*}}{(m_B-m_{D^*})^2}
\right) \xi^2(w) \label{dgdwds}
\end{eqnarray}
In the strict SV limit we have
\begin{equation}\label{svlim}
m_B=m_b\qquad m_{D^*}=m_D=m_c\qquad m_c =m_b(1-\epsilon)
\end{equation}
where $\epsilon\equiv (m_b-m_c)/m_b$ is a small parameter.

The variable $w$ is related to the dilepton invariant mass
$q^2$ through
\begin{equation}\label{q2w}
q^2 = m^2_B + m^2_D- 2 m_B m_D w
\end{equation}
The kinematic limits of $q^2$ are easily identified as
\begin{equation}\label{q2minmax}
q^2_{max}=(m_b - m_c)^2\qquad q^2_{min}=0
\end{equation}
The corresponding limits of $w$ are
\begin{equation}\label{wminmax}
w_{min}=1\qquad w_{max}=\frac{m^2_b+m^2_c}{2 m_b m_c}
\end{equation}
Defining $s\equiv w-1$ we have $0\leq s\leq \epsilon^2/2$,
where the upper limit is valid to leading order in $\epsilon$.
Expanded to leading order in $\epsilon$, (\ref{dgdwd}) gives
\begin{equation}\label{gamd}
\Gamma(B\to D l\nu)
= \frac{G^2_F |V_{cb}|^2}{6\pi^3}m^5_b \sqrt{2}
\int^{\epsilon^2/2}_{0} s^{3/2}\, ds =
\frac{G^2_F |V_{cb}|^2}{60\pi^3}(m_b-m_c)^5
\end{equation}
which is the decay rate in the SV limit. In this derivation
we have made use of the fact that $\xi(w)=\xi(1)+{\cal O}(\epsilon^2)$,
which can be approximated by $\xi(1)=1$. In this way any dependence
on nontrivial hadronic input has disappeared.
Similarly we can expand the integral over (\ref{dgdwds}) in
$\epsilon$ to extract the leading contribution in the SV limit.
We obtain
\begin{equation}\label{gamds}
\Gamma(B\to D^* l\nu)
= \frac{G^2_F |V_{cb}|^2}{20\pi^3}(m_b-m_c)^5
\end{equation}
We also observe that higher $D$-meson resonances and 
hadronic multiparticle states have wave functions of
the light degrees of freedom that are orthogonal to the
ground state wave function of the light cloud (identical
for $D$, $D^*$ and $B$) in the SV limit. There is therefore 
no overlap of those higher excitations with the initial $B$
and the corresponding rates vanish.

Finally, we need to take the SV limit of the inclusive rate
as obtained from the heavy quark expansion in (\ref{bclnu}).
In this limit the second-order power corrections and perturbative QCD
corrections disappear, and we only have to expand the phase space function
$z_3$ in the small-$\epsilon$ limit. We find 
$z_3=64 \epsilon^5/5 +{\cal O}(\epsilon^6)$ and 
\begin{equation}\label{gamxc}
\Gamma(B\to X_c l\nu)
= \frac{G^2_F |V_{cb}|^2}{15\pi^3}(m_b-m_c)^5
\end{equation}
We see that indeed the inclusive HQE result (\ref{gamxc}) is saturated
by the sum of just the two exclusive rates (\ref{gamd}) and (\ref{gamds}).
Clearly, the SV limit is a very special situation.
Nevertheless, it is an interesting example of exact (local) quark-hadron
duality. Moreover, the semileptonic rates into $D$ and $D^*$
measured in experiment account for roughly two thirds of the
inclusive rate, indicating that the SV limit is not even entirely
unrealistic.

\section{QCD factorization in exclusive hadronic $B$ decays}

\subsection{Introduction}

Decay amplitudes for exclusive nonleptonic $B$ decays,
such as $B\to\pi\pi$,
can be computed starting from the effective weak
Hamiltonian discussed in sec. 3.3.
Whereas the Wilson coefficients $C_i$ are well understood,
the main problem is posed by the hadronic matrix elements
of the operators $Q_i$.
In some cases this problem can be circumvented
(CP asymmetry in $B\to J/\Psi K_S$), or at least reduced
using SU(2) or SU(3) flavour symmetries and an appropriate
combination of various channels.
However, an improved understanding of the QCD dynamics in 
exclusive hadronic
$B$ decays would greatly enhance our capability to extract
from these processes the underlying flavour physics.

Indeed, it turns out that
the heavy-quark limit leads to substantial simplifications
also in the problem of hadronic two-body decays of heavy
hadrons. Again the main feature is the factorization of
short-distance and long-distance contributions.
In the case of the matrix elements of four-quark operators
$Q_i$ the factorization takes the form
\begin{eqnarray}\label{fform}
&&\langle\pi(p')\pi(q)|Q_i|\bar B(p)\rangle =
f^{B\to\pi}(q^2)\int^1_0 du\, T^{I}_i(u)\Phi_\pi(u)
 \nonumber \\
&& \ \ +\int^1_0 d\xi du dv\, T^{II}_i(\xi,u,v)\Phi_B(\xi)
\Phi_\pi(u) \Phi_\pi(v)
\end{eqnarray}
This {\it factorization formula\/} is valid up to corrections of 
relative order $\Lambda_{QCD}/m_b$.
Here $f^{B\to\pi}(q^2)$ is a $B\to\pi$ form factor
evaluated at $q^2=m^2_\pi\approx 0$, and $\Phi_\pi$ ($\Phi_B$) are 
leading-twist light-cone
distribution amplitudes {``wave functions'') of the pion ($B$ meson).
These objects contain the long-distance dynamics.
The short-distance physics, dominated by scales of order $m_b$, 
is described by the 
hard-scattering kernels $T^{I,II}_i$, 
which are calculable in perturbation theory. $T^{I}_i$ starts 
at ${\cal O}(\alpha^0_s)$, $T^{II}_i$ at ${\cal O}(\alpha^1_s)$
(see Fig.~\ref{figlett1}). 
\begin{figure}
[htbp]
\centering
\vspace*{-2cm}
\includegraphics[width=20cm,height=28cm]{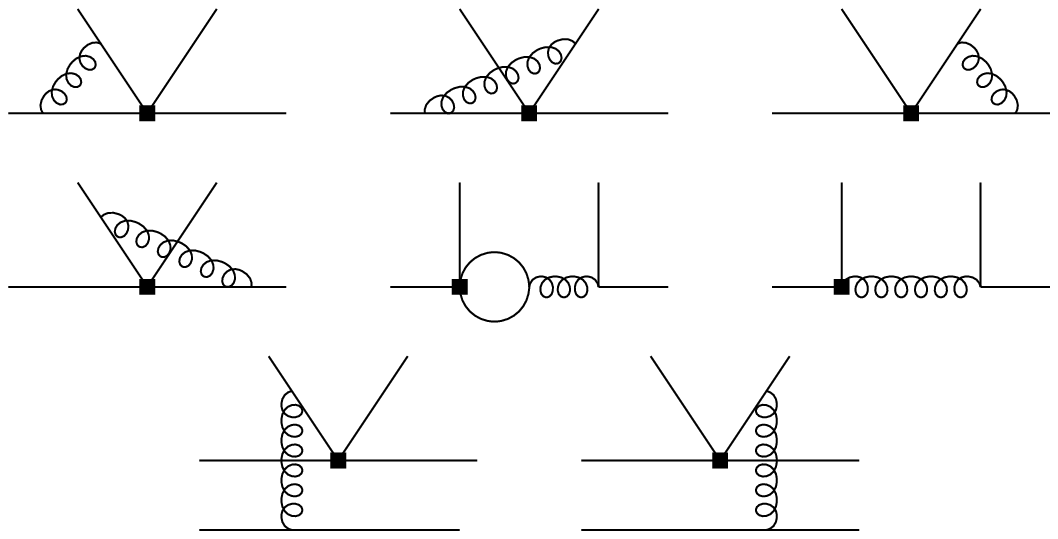}
\vspace*{-19cm}
\caption{Order $\alpha_s$ corrections to the hard 
scattering kernels $T^I_i$ (first two rows) and $T^{II}_i$ 
(last row). In the case of $T^I_i$, the spectator quark does 
not participate in the hard interaction and is not drawn. 
The two lines directed upwards represent the two quarks forming
the emitted pion.}
\label{figlett1}
\end{figure}
In (\ref{fform}) long- and short-distance contributions
are thus systematically disentangled, that is {\it factorized}.
The long-distance sensitive quantities (form factors and
wave functions) still need to be determined by other means, but they 
are universal quantities and much simpler than the original
full $B\to\pi\pi$ matrix elements we started with.
They could in principle be calculated by nonperturbative methods
or extracted experimentally from other observables.
In any case (\ref{fform}) represents a substantial
simplification of our problem.   

The general expression (\ref{fform}) further simplifies
when we neglect perturbative $\alpha_s$-corrections.
The $T^{II}$ term is then absent and the kernel $T^I$
becomes a constant in $u$, such that the pion distribution
amplitude integrates to the pion decay constant.
The matrix element of operator $Q^u_1$, for instance,
reduces to
\begin{equation}\label{nfact}
\langle\pi^+\pi^-|(\bar ub)_{V-A}(\bar du)_{V-A}|\bar B\rangle
\to\langle\pi^+|(\bar ub)_{V-A}|\bar B\rangle\cdot
 \langle\pi^-|(\bar du)_{V-A}|0\rangle
= i m^2_B f^{B\to\pi}(0) f_\pi
\end{equation}
This procedure, termed ``naive factorization'' 
has long been used in phenomenological
application, but the justification had been less clear. 
An obvious issue is the
scheme and scale dependence of the matrix elements of four-quark
operators, which is needed to cancel the corresponding dependence
in the Wilson coefficients. This dependence is lost
in naive factorization as the factorized currents are scheme
independent objects. In QCD factorization (\ref{fform})
the proper scale and scheme dependence is recovered by the
inclusion of ${\cal O}(\alpha_s)$ corrections as we will
see explicitly below.

A qualitative justification for (\ref{nfact}) had been
given by Bjorken (1989). It is based on the {\it colour
transparency\/} of the hadronic environment for the highly energetic
pion emitted in $B$ decay (the $\pi^-$ in the
above example, which is being created from the vacuum).
This is related to the decoupling of soft gluons from
the small-size colour-singlet object that the emitted pion represents.
The QCD factorization approach as encoded in (\ref{fform})
may be viewed as a consistent formalization and generalization
of  Bjorken's colour transparency argument.
This treatment of hadronic $B$ decays is based on the analysis of
Feynman diagrams in the heavy quark limit, utilizing consistent
power counting to identify the leading contributions. The framework
is very similar in spirit to more conventional applications  of
perturbative QCD in exclusive hadronic processes with a large 
momentum transfer, as the pion electromagnetic form factor 
(see the article by Sterman and Stoler (1997) for a recent review).
It justifies and extends the ansatz of naive factorization. In particular
the method includes, for $B\to\pi\pi$, the hard
nonfactorizable spectator interactions, penguin contributions
and rescattering effects (Fig.~\ref{figlett1}). 
As a corollary, one finds that strong
rescattering phases are either of ${\cal O}(\alpha_s)$, and
calculable, or power suppressed. In any case they vanish therefore
in the heavy quark limit.
QCD factorization is valid for cases where the emitted particle
(the meson created from the vacuum in the weak process, as opposed
to the one that absorbs the $b$-quark spectator) is a small size
colour-singlet object, e.g. either a fast light meson
($\pi$, $\varrho$, $K$, $K^*$) or a $J/\Psi$.

Note that the term {\it factorization\/} is used here for
two a priori entirely different things: In the case
of QCD factorization (\ref{fform}), it refers to
the factorization of short-distance and long-distance contributions.
In the sense of the phenomenological approach of naive
factorization (\ref{nfact}), it simply denotes the 
separation of the hadronic matrix element of a four-quark
operator into two factors of matrix elements of bilinear
currents. It is a nontrivial result that the latter, naive
factorization is obtained as the lowest order approximation
of QCD factorization.
To avoid confusion, it is useful to keep the distinction
in mind. For example, the hard gluon exchange corrections
between the two quark currents in Fig. \ref{figlett1}
are ``nonfactorizable'' in the sense of naive factorization,
although they are a consistent ingredient of (\ref{fform}),
hence ``factorizable'' in the sense of QCD.

In the following we shall
discuss QCD factorization in some detail
using the example of $B\to D\pi$ decays.
In this case the $b\to u$ transition current is
replaced by a heavy-heavy $b\to c$ current. This
case is somewhat simpler than $B\to \pi\pi$ since the
spectator interaction (the $T^{II}$ term, 
bottom line of Fig. \ref{figlett1})
does not contribute to leading power. This is because for
a heavy-to-heavy transition the spectator quark, and hence the 
gluon attached to it, is always soft.
This leads to a suppression, according to the colour transparency argument,
when this gluon couples to the emitted pion.
Also penguin contributions are absent for $B\to D\pi$.
We shall illustrate explicitly how factorization emerges at the one-loop 
order in this specific case, and in the heavy-quark limit, defined as
$m_b$, $m_c\gg\Lambda_{\rm QCD}$ with $m_c/m_b$ fixed.

Further details on QCD factorization in $B$ decays
and additional literature can be found in the articles by
Beneke et al. (1999, 2000 \& 2001).

\newpage

\subsection{$B\to D\pi$: Factorization to one-loop order}

\subsubsection{Preliminaries}

The effective Hamiltonian relevant for $B\to D\pi$ can be written as
\begin{equation}\label{heff18}
{\cal H}_{\rm eff}=\frac{G_F}{\sqrt{2}}V^*_{ud}V_{cb}
\left( C_0 O_0+C_8 O_8\right) ,
\end{equation}
with the operators
\begin{eqnarray}\label{o18}
O_0 &=& \bar c\gamma^\mu(1-\gamma_5)b\, 
        \bar d\gamma_\mu(1-\gamma_5)u , \\
O_8 &=& \bar c\gamma^\mu(1-\gamma_5)T^a b\, 
        \bar d\gamma_\mu(1-\gamma_5)T^a u .
\end{eqnarray}
Here we have chosen to write the two independent operators
in the singlet-octet basis, which is most convenient for
our purposes, rather than in the more conventional bases
of $Q_1$, $Q_2$ or $Q_+$, $Q_-$ 
(see the discussion in sec. 3; because all four quark flavours
are different in (\ref{heff18}), penguin operators are absent). 
The Wilson coefficients $C_0$, $C_8$ 
have been calculated at next-to-leading order
in renormalization-group improved perturbation theory
(Altarelli et al. 1981, Buras \& Weisz 1990) and are given by
\begin{equation}\label{c18}
C_0=\frac{N_c+1}{2N_c}C_++\frac{N_c-1}{2N_c}C_- ,\qquad
C_8=C_+-C_- ,
\end{equation}
where
\begin{equation}\label{cpm}
C_\pm(\mu)=\left(1+\frac{\alpha_s(\mu)}{4\pi}B_\pm\right)\,
\bar C_\pm(\mu) ,
\end{equation}
\begin{equation}\label{cpmb}
\bar C_\pm(\mu)=\left[\frac{\alpha_s(M_W)}{\alpha_s(\mu)}\right]^{d_\pm}
\left[1+\frac{\alpha_s(M_W)-\alpha_s(\mu)}{4\pi}(B_\pm-J_\pm)\right] .
\end{equation}
(The coefficients $C_0$, $C_8$ are related
to the ones of the standard basis by $C_0=C_1+C_2/3$ and 
$C_8=2C_2$.) 
We employ the next-to-leading order expression for the running 
coupling,
\begin{equation}\label{als}
\alpha_s(\mu)=\frac{4\pi}{\beta_0\ln(\mu^2/\Lambda_{\rm QCD}^2)}
 \left[1-\frac{\beta_1}{\beta^2_0}
  \frac{\ln\ln(\mu^2/\Lambda_{\rm QCD}^2)}{\ln(\mu^2/\Lambda_{\rm
      QCD}^2)} \right] ,
\end{equation}
\begin{equation}\label{b0b1}
\beta_0=\frac{11N_c-2f}{3} ,\qquad 
\beta_1=\frac{34}{3}N^2_c-\frac{10}{3}N_c f-2 C_F f ,\qquad
C_F=\frac{N^2_c-1}{2N_c} ,
\end{equation}
where $N_c$ is the number of colours, and $f$ the number of light 
flavours. $\Lambda_{\rm QCD}\equiv\Lambda^{(f)}_{\overline{\rm MS}}$ is 
the QCD scale in the $\overline{\rm MS}$ scheme with $f$ flavours.
Next we have
\begin{equation}\label{dgbpm}
d_\pm=\frac{\gamma^{(0)}_\pm}{2\beta_0} ,\qquad
\gamma^{(0)}_\pm=\pm 12\frac{N_c\mp 1}{2N_c} ,\qquad
B_\pm=\pm\frac{N_c\mp 1}{2N_c}B .
\end{equation}
The general definition of $J_\pm$ may be found in 
(Buchalla, Buras \& Lautenbacher 1996).
Numerically, for $N_c=3$ and $f=5$
\begin{equation}\label{dbjpm}
d_\pm=\left\{ 
 \begin{array}{c}
    \phantom{-}\frac{6}{23} , \\[0.2cm]
    -\frac{12}{23} ,
 \end{array} \right.
\qquad
B_\pm-J_\pm=\left\{ 
 \begin{array}{c}
   \phantom{-}\frac{6473}{3174} , \\[0.2cm]
   -\frac{9371}{1587} .
 \end{array} \right.
\end{equation}
The quantities $\beta_0$, $\beta_1$, $d_\pm$, $B_\pm-J_\pm$
are scheme independent. The scheme dependence of the coefficients
at next-to-leading order is parametrized by $B_\pm$ in (\ref{cpm}). In 
the naive dimensional regularization (NDR) and `t~Hooft-Veltman (HV)
schemes, this scheme dependence is expressed in a single number
$B$ with $B_{\rm NDR}=11$ and $B_{\rm HV}=7$.
The dependence of the Wilson coefficients on the renormalization
scheme and scale is cancelled by a corresponding scale and scheme dependence 
of the hadronic matrix elements of the operators $O_0$ and $O_8$.

Before continuing with a discussion 
of these matrix elements, it is useful to consider the
flavour structure for the various contributions to $B\to D\pi$ decays.  
The possible quark-level topologies are depicted in 
Fig.~\ref{fig:bdpi}. 
\begin{figure}
 \vspace{4cm}
\includegraphics{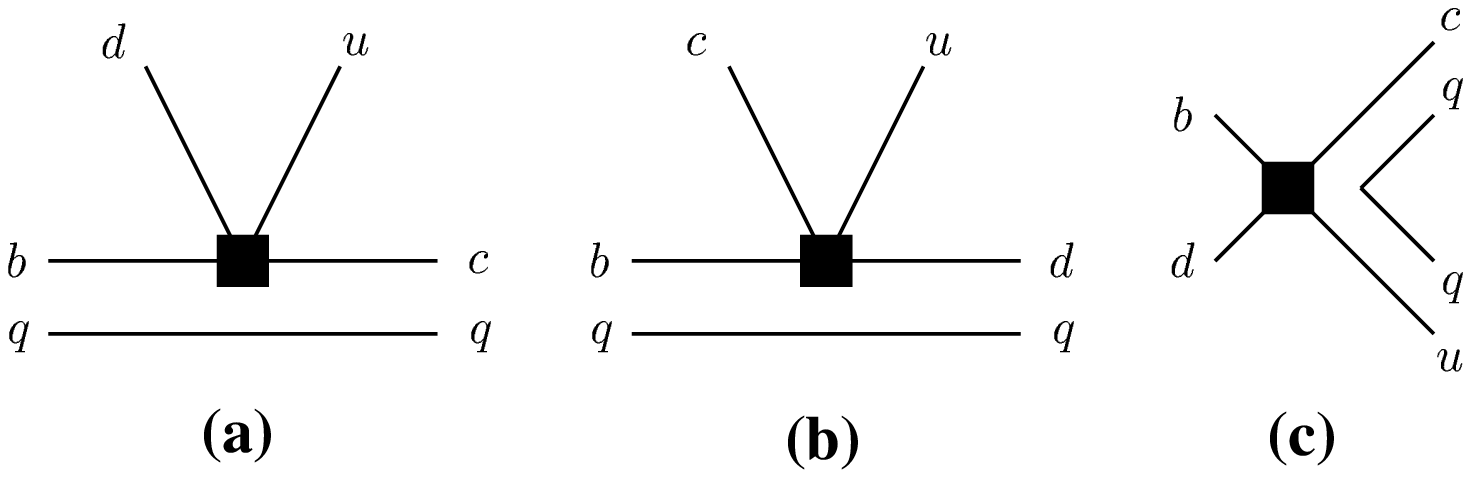}
 \caption{\small Basic quark-level topologies for $B\to D\pi$
   decays ($q=u$, $d$): (a) class-I, (b) class-II,  
   (c) weak annihilation. $\bar B_d\to D^+\pi^-$ receives 
   contributions from (a) and (c), $\bar B_d\to D^0\pi^0$ from
   (b) and (c), and $B^-\to D^0\pi^-$ from (a) and (b).
   Only (a) contributes in the heavy-quark limit.
    \label{fig:bdpi}}
\end{figure}
In the terminology generally adopted for
two-body non-leptonic decays, the decays $\bar B_d\to D^+\pi^-$,
$\bar B_d\to D^0\pi^0$ and $B^-\to D^0\pi^-$ are referred to
as class-I, class-II and class-III, respectively.
In both $\bar B_d\to D^+\pi^-$ and $B^-\to D^0\pi^-$ decays
the pion can be directly created from the weak current.
We may call this a class-I contribution, following the above
terminology. In addition, in the case of $\bar B_d\to D^+\pi^-$
there is a contribution from weak annihilation and a class-II
amplitude contributes to $B^-\to D^0\pi^-$, see Fig.~\ref{fig:bdpi}. 
The important point is that the spectator quark goes into the 
light meson in the case of the class-II amplitude. 
This amplitude is suppressed in the 
heavy-quark limit, as is the annihilation amplitude. 
The amplitude for $\bar B_d\to D^0\pi^0$, receiving
only class-II and annihilation contributions, is therefore
subleading compared
with $\bar B_d\to D^+\pi^-$ and $B^-\to D^0\pi^-$, which are
dominated by the class-I topology.
The treatment of this leading class-I mechanism will be the
main subject of the following sections. 
We shall use the one-loop analysis for $\bar B_d\to D^+\pi^-$ as a
concrete example on which we will illustrate explicitly
how the factorization formula can be derived.

\subsubsection{Soft and collinear cancellations at one-loop}

In order to demonstrate the property of factorization for
$\bar B_d\to D^+\pi^-$, we have to analyze the ``non-factorizable''
one-gluon exchange contributions (Fig.~\ref{fig6}) 
\begin{figure}
[htbp]
\centering
\vspace*{-4cm}
\includegraphics[width=18cm,height=27cm]{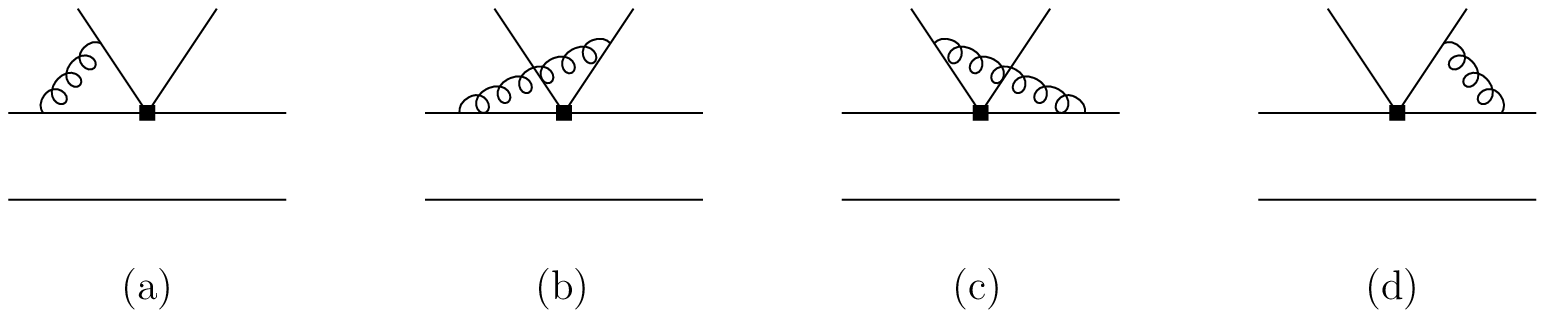}
\vspace*{-20cm}
\caption{``Non-factorizable'' vertex corrections.}
\label{fig6}
\end{figure}
to the $b\to c\bar ud$ transition. 
We consider the leading, valence Fock state
of the emitted pion. This is justified since higher Fock components
only give power-suppressed contributions to the decay
amplitude in the heavy-quark limit.
The valence Fock state of the pion can be written as
\begin{equation}\label{piwf}
|\pi(q)\rangle =\int\frac{du}{\sqrt{u\bar u}}
\frac{d^2 l_\perp}{16\pi^3}\frac{1}{\sqrt{2N_c}}
\left(a^\dagger_\uparrow(l_q)b^\dagger_\downarrow(l_{\bar q})-
      a^\dagger_\downarrow(l_q)b^\dagger_\uparrow(l_{\bar q})\right)
|0\rangle\, \Psi(u,\vec l_\perp) ,
\end{equation}
where $a^\dagger_s$ ($b^\dagger_s$) denotes the creation operator
for a quark (antiquark) in a state with spin $s=\uparrow$ or
$s=\downarrow$, and we have suppressed colour indices. This 
representation of the pion state is adequate for a leading-power 
analysis. The wave function $\Psi(u,\vec l_\perp)$ is defined as the
amplitude for the pion to be composed of two on-shell quarks,
characterized by longitudinal momentum fraction $u$ and
transverse momentum $l_\perp$. 
The on-shell momenta
($l^2_{q,\bar q}=0$) of the quark ($l_q$) and the antiquark ($l_{\bar q}$)
are given by
\begin{equation}\label{q1u}
l_q=u q + l_\perp+\frac{\vec{l}^{\,2}_\perp}{4 u E}n_- ,
\qquad
l_{\bar q}=\bar u q - l_\perp+\frac{\vec{l}^{\,2}_\perp}{4\bar u E}n_- .
\end{equation}
Here $q=E(1,0,0,1)$ is the pion momentum, $E=p_B\cdot q/m_B$
the pion energy and $n_-=(1,0,0,-1)$. 
Furthermore $l_\perp\cdot q=l_\perp\cdot n_-=0$.
For the purpose of power counting 
$l_\perp\sim\Lambda_{\rm QCD}\ll E\sim m_b$.
Note that the invariant mass of the valence state is
$(l_q+l_{\bar q})^2=\vec{l}^{\,2}_\perp/(u\bar u)$, which is of order
$\Lambda^2_{\rm QCD}$ and hence negligible in the heavy-quark limit,
unless $u$ is in the vicinity of the endpoints ($0$ or $1$).
In this case the invariant mass of the quark-antiquark pair
becomes large and the valence Fock state is no longer
a valid representation of the pion. However, in the heavy-quark
limit the dominant contributions to the decay amplitude come from
configurations where both partons are hard
($u$ and $\bar u$ both of order 1) and the two-particle Fock state
yields a consistent description. The suppression of the
soft regions ($u$ or $\bar u\ll 1$) is related to the endpoint
behaviour of the pion wave function. 
We will provide an explicit consistency
check of this important feature later on.

As a next step we write down the amplitude
\begin{equation}\label{piudb}
\langle\pi(q)|u(0)_\alpha\bar d(y)_\beta|0\rangle =
\int du\frac{d^2l_\perp}{16\pi^3}\frac{1}{\sqrt{2 N_c}}
\Psi^*(u,\vec l_\perp)(\gamma_5\!\not\! q)_{\alpha\beta}\,
 e^{i l_q\cdot y} ,
\end{equation}
which appears as an ingredient of the $B\to D\pi$ matrix element.
The right-hand side of (\ref{piudb}) follows directly from (\ref{piwf}).
Using (\ref{piudb}) it is straightforward to write down the
one-gluon exchange contribution to the $B\to D\pi$ matrix
element of the operator $O_8$ (Fig.~\ref{fig6}). We have
\begin{eqnarray}\label{o8a1a2}
\langle D^+\pi^-|O_8|\bar B_d\rangle_{\rm 1-gluon} &=& 
\\
&&\hspace*{-4cm}
i g_s^2\frac{C_F}{2}\int\frac{d^4k}{(2\pi)^4}
\langle D^+|\bar c A_1(k) b|\bar B_d\rangle\frac{1}{k^2}
\int^1_0 du\frac{d^2l_\perp}{16\pi^3}
\frac{\Psi^*(u,\vec l_\perp)}{\sqrt{2 N_c}}\,
{\rm tr}[\gamma_5\!\not\! q A_2(l_q,l_{\bar q},k)] , \nonumber
\end{eqnarray}
where
\begin{equation}\label{a1bc}
A_1(k)=
\frac{\gamma^\lambda(\not\! p_c-\not\! k+m_c)\Gamma}{2p_c\cdot k-k^2}-
\frac{\Gamma(\not\! p_b+\not\! k+m_b)\gamma^\lambda}{2p_b\cdot k+k^2} ,
\end{equation}
\begin{equation}\label{a2ud}
A_2(l_q,l_{\bar{q}},k)=
\frac{\Gamma(\not\! l_{\bar q}+\not\! k)\gamma_\lambda}{2l_{\bar q}\cdot k+
      k^2}-
\frac{\gamma_\lambda(\not\! l_q+\not\! k)\Gamma}{2l_q\cdot k+k^2} .
\end{equation}
Here $\Gamma=\gamma^\mu(1-\gamma_5)$ and $p_b$, $p_c$ are the
momenta of the $b$ quark and the $c$ quark, respectively.
Note that this expression holds in an arbitrary covariant gauge. 
The gauge-parameter dependent part of the gluon propagator gives no
contribution to (\ref{o8a1a2}), as can be seen
from (\ref{a1bc}) and (\ref{a2ud}). There is no correction 
to the matrix element of $O_0$ at order $\alpha_s$, because in this 
case the $(d\bar{u})$ pair is necessarily in a colour-octet 
configuration and cannot form a pion.

In (\ref{o8a1a2}) the pion wave function $\Psi(u,l_\perp)$ 
appears separated
from the $B\to D$ transition. This is merely
a reflection of the fact that we have represented the pion state 
by (\ref{piwf}). It does not, by itself,
imply factorization, since the right-hand side 
of (\ref{o8a1a2}) involves
still nontrivial integrations over $\vec l_\perp$ and gluon
momentum $k$, and long- and short-distance contributions are not
yet disentangled. In order for (\ref{o8a1a2}) to make sense 
we need to show that the integral over $k$ receives only subdominant 
contributions from the region of small $k^2$. This is equivalent 
to showing that the integral over $k$ does not contain 
infrared divergences at leading power in $1/m_b$. 

To demonstrate infrared finiteness of the one-loop integral
\begin{equation}
 \label{dka1a2}
J\equiv
\int d^4k\, \frac{1}{k^2}\, A_1(k)\otimes A_2(l_q,l_{\bar{q}},k)
\end{equation}
at leading power, the heavy-quark limit and the
corresponding large light-cone momentum of the pion
are again essential. First note that when $k$ is of order $m_b$, 
$J\sim 1$ for dimensional reasons. Potential infrared divergences 
could arise when $k$ is soft or when $k$ is collinear to the 
pion momentum $q$. We need to show that the contributions from 
these regions are power suppressed in $m_b$. (Note that we do not 
need to show that $J$ is infrared finite. It is enough that 
logarithmic divergences have coefficients that are 
power suppressed.)

We treat the soft region of integration first. 
Here all components of $k$ become
small simultaneously, which we describe by scaling $k\sim\lambda$.
Counting powers of $\lambda$
($d^4k\sim \lambda^4$, $1/k^2\sim\lambda^{-2}$,
$1/p\cdot k\sim\lambda^{-1}$) reveals that each of the four diagrams
(corresponding to the four terms in the product in
(\ref{dka1a2})) is logarithmically divergent. However, because 
$k$ is small, the integrand can be simplified. For instance, 
the second term in $A_2$ can be approximated as
\begin{equation}
\label{a2lperp}
\frac{\gamma_\lambda(\not\! l_q+\not\! k)\Gamma}{2 l_q\cdot k+k^2}=
\frac{\gamma_\lambda(\!u\not\! q+\not\! l_\perp+
  \frac{\vec l^2_\perp}{4uE}\not\! n_-+\not\! k)\Gamma}{2u q\cdot k+
   2l_\perp\cdot k+\frac{\vec l^2_\perp}{2uE} n_-\cdot k+k^2}
   \simeq \frac{q_\lambda}{q\cdot k}\,\Gamma ,
\end{equation}
where we used that $\not\!q$ to the extreme left or right of an 
expression gives zero due to the on-shell condition for the external 
quark lines. We get exactly the same expression but with an opposite 
sign from the other term in $A_2$ and hence the soft 
divergence cancels out when adding the two terms in $A_2$. More 
precisely, we find that the integral is infrared finite in the 
soft region when $l_\perp$ is neglected. When $l_\perp$ is not 
neglected, there is a divergence from soft $k$ which is proportional 
to $l^2_\perp/m_b^2\sim \Lambda^2_{\rm QCD}/m_b^2$. In either case, the 
soft contribution to $J$ is of order $\Lambda_{\rm QCD}/m_b$ 
or smaller and 
hence suppressed relative to the hard contribution.
This corresponds to the standard soft cancellation mechanism,
which is a technical manifestation of colour transparency.

Each of the four terms in (\ref{dka1a2}) is also divergent 
when $k$ becomes collinear with the light-cone
momentum $q$. 
It is convenient to introduce, for any four-vector $v$, the
light-cone components
\begin{equation}\label{vpmv03}
v_\pm=\frac{v^0\pm v^3}{\sqrt{2}}
\end{equation}
The collinear region is then described by
the scaling 
\begin{equation}\label{kcoll}
k^+\sim \lambda^0,\quad k_\perp\sim\lambda,\quad k^-\sim\lambda^2 .
\end{equation}
Then $d^4k\sim dk^+ dk^- d^2k_\perp\sim\lambda^4$ and
$q\cdot k=q^+ k^-\sim\lambda^2$, $k^2=2k^+ k^-+k^2_\perp\sim\lambda^2$.
The divergence is again logarithmic and it is thus sufficient
to consider the leading behaviour in the collinear limit. 
Writing $k=\alpha q+\ldots$ we can now simplify the second term 
of $A_2$ as
\begin{equation}
\label{a2lperp2}
\frac{\gamma_\lambda(\not\! l_q+\not\! k)\Gamma}{2 l_q\cdot k+k^2}
\simeq 
q_\lambda\,\frac{2 (u+\alpha)\Gamma}{2 l_q\cdot k+k^2} .
\end{equation}
No simplification occurs in the denominator (in particular $l_\perp$ 
cannot be neglected), but the important point is that the leading-power 
contribution is proportional to $q_\lambda$. Therefore, 
substituting $k=\alpha q$ into $A_1$ and using $q^2=0$, we obtain 
\begin{equation}\label{a1coll}
q_\lambda A_1\simeq
\frac{\not\! q(\not\! p_c+m_c)\Gamma}{2\alpha p_c\cdot q}-
\frac{\Gamma(\not\! p_b+m_b)\!\not\! q}{2\alpha p_b\cdot q}=0 ,
\end{equation}
employing the equations of motion for the heavy quarks.
Hence the collinearly 
divergent region is seen to cancel out via the standard
collinear Ward identity.
This completes the proof of the absence of infrared divergences 
at leading power in the hard-scattering kernel for 
$\bar B_d\to D^+\pi^-$ to one-loop order. In other words, we have shown 
that the ``non-factorizable'' diagrams of Fig.~\ref{fig6} are 
dominated by hard gluon exchange. 
The proof at two loops has been given by Beneke et al. (2000)
and a proof to all orders by Bauer et al. (2001c).

Since we have now established that the leading contribution to 
$J$ arises from $k$ of order $m_b$ (``hard'' $k$), and 
since $|\vec l_\perp|\ll E$, we may expand $A_2$
in $|\vec l_\perp|/E$. To leading power the expansion simply 
reduces to neglecting $l_\perp$ altogether, which implies
$l_q=uq$ and $l_{\bar q}=\bar uq$ in (\ref{a2ud}). 
As a consequence,
we may perform the $l_\perp$ integration in (\ref{o8a1a2})
over the pion wave function. Defining
\begin{equation}\label{psiphi}
\int\frac{d^2l_\perp}{16\pi^3}
\frac{\Psi^*(u,\vec l_\perp)}{\sqrt{2N_c}}\equiv
\frac{i f_\pi}{4 N_c}\Phi_\pi(u) , 
\end{equation}
the matrix element of $O_8$ in (\ref{o8a1a2}) becomes
\begin{eqnarray}\label{o8phi}
\langle D^+\pi^-|O_8|\bar B_d\rangle_{\rm 1-gluon} &=& \\
&&\hspace*{-4cm}-g_s^2\frac{C_F}{8N_c}\int\frac{d^4k}{(2\pi)^4}
\langle D^+|\bar c A_1(k) b|\bar B_d\rangle\frac{1}{k^2} f_\pi
\int^1_0 du\, \Phi_\pi(u)\,
{\rm tr}[\gamma_5\!\not\! q A_2(uq,\bar uq,k)] . \nonumber
\end{eqnarray}
Putting $y$ on the light-cone in (\ref{piudb}),
$y^+=y_\perp=0$, hence $l_q\cdot y=l^+_q y^-=uq y$, 
we see that the $l_\perp$-integrated wave
function $\Phi_\pi(u)$ in (\ref{psiphi}) is precisely the
light-cone distribution amplitude of the pion in the usual
definition. Indeed,
the leading-twist light-cone distribution amplitude for pseudoscalar mesons 
($P$) reads 
\begin{equation}
\label{distamps}
\langle P(q)|\bar{q}(y)_{\alpha} q'(x)_{\beta}|0\rangle\Big|_{(x-y)^2=0}
= \frac{i f_P}{4}\,(\not\!q\gamma_5)_{\beta
\alpha}\,\int_0^1 du\,e^{i(\bar{u} qx+u qy)}\,\Phi_P(u,\mu)
\end{equation}
Here it is understood that the operator on the 
left-hand side is a colour singlet. We use the 
``bar''-notation, i.e.\ $\bar{v}\equiv 1-v$ for 
any longitudinal momentum fraction variable. The parameter 
$\mu$ is the renormalization scale of the 
light-cone operators on the left-hand side. The distribution amplitude 
is normalized as $\int_0^1 du\,\Phi_{P}(u,\mu)=1$. 
One defines the asymptotic distribution amplitude as the limit in 
which the renormalization scale is sent to infinity. 
The asymptotic form is
\begin{equation}
\label{asform}
\Phi_{P}(u,\mu)\stackrel{\mu\to\infty}{=} 6 u\bar{u}.
\end{equation}
The decay constant appearing in (\ref{distamps}) refers to the normalization 
in which $f_\pi=131\,$MeV. There is a 
path-ordered exponential that connects the two quark fields at 
different positions and makes the light-cone operators gauge invariant. 
In (\ref{distamps}) we have suppressed this standard factor.

This derivation demonstrates the relevance of the 
light-cone wave function for the factorization formula. Note 
that the collinear approximation for the quark-antiquark momenta 
emerges automatically in the heavy-quark limit. 

After the $k$ integral is performed, the expression (\ref{o8phi}) 
can be cast into the form 
\begin{equation}\label{o8fth}
\langle D^+\pi^-|O_8|\bar B_d\rangle_{\rm 1-gluon} \sim  
F_{B\to D}(0)\, \int^1_0 du\, T_8(u,z) \Phi_\pi(u) , 
\end{equation}
where $z=m_c/m_b$, $T_8(u,z)$ is the hard-scattering kernel, and 
$F_{B\to D}(0)$ the form factor that parametrizes the 
$\langle D^+|\bar c [\ldots] b|\bar B_d\rangle$ matrix element. 
The result for $T_8(u,z)$ will be given in the following
section.

\subsubsection{Matrix elements at next-to-leading order}

As we have seen above, the $\bar B_d\to D^+\pi^-$ amplitude
factorizes in the heavy-quark limit into a matrix element of the 
form $\langle D^+|\bar{c}[\ldots]b|\bar B_d\rangle$ for the 
$B\to D$ transition and a 
matrix element $\langle\pi^-|\bar{d}(x)[\ldots]u(0)|0\rangle$ 
with $x^2=0$ that gives rise to the pion light-cone distribution 
amplitude. Leaving aside power-suppressed contributions, the essential
requirement for this conclusion was the absence of both soft
and collinear infrared divergences in the gluon exchange between
the $(\bar cb)$ and $(\bar du)$ currents.
This gluon exchange is therefore calculable in QCD perturbation
theory. We now present these corrections explicitly to order 
$\alpha_s$.

The effective Hamiltonian (\ref{heff18}) can be written as
\begin{eqnarray}\label{bheff}
   {\cal H}_{\rm eff} &=& \frac{G_F}{\sqrt2}\,V^*_{ud} V_{cb}\,
    \Bigg\{ \left[ \frac{N_c+1}{2N_c}\bar C_+(\mu)
    + \frac{N_c-1}{2N_c}\bar C_-(\mu)
    + \frac{\alpha_s(\mu)}{4\pi}\,\frac{C_F}{2N_c}\,B C_8(\mu)
    \right] O_0 \nonumber\\
   &&\qquad \mbox{}+ C_8(\mu)\,O_8 \Bigg\} ,
\end{eqnarray}
where the scheme-dependent terms in the coefficient of the operator 
$O_0$, proportional to the constant $B$ 
defined after (\ref{dbjpm}), have been written explicitly.

Schematically, the matrix elements of both $O_0$ and $O_8$ can be
expressed in a form analogous to (\ref{o8fth}). Because the light-quark
pair has to be in a colour singlet to produce the pion in the leading
Fock state, only $O_0$ gives a contribution to zeroth order in
$\alpha_s$. Similarly, to first order in $\alpha_s$ only $O_8$
can contribute. The result of 
computing the diagrams in Fig.~\ref{fig6} with an insertion of $O_8$ can 
be written in a form that holds simultaneously for $H=D,D^*$ and 
$L=\pi, \rho$, using only that the $(\bar{u}d)$ pair is a colour 
singlet and that the external quarks can be taken on-shell.
One obtains ($z=m_c/m_b$)
\begin{eqnarray}\label{delo8}
   \langle H(p') L(q)|O_8|\bar{B}_d(p)\rangle
   &=& \frac{\alpha_s}{4\pi}\frac{C_F}{2N_c}\,i f_L \int_0^1 du\,
   \Phi_L(u) \\ 
   &&\hspace{-4cm}\times 
    \left[ - \left( 6\ln\frac{\mu^2}{m_b^2} + B \right)
    (\langle J_V\rangle - \langle J_A\rangle)
    + F(u,z)\,\langle J_V\rangle - F(u,-z)\,\langle J_A\rangle \right] ,
    \nonumber
\end{eqnarray}
where 
\begin{equation}\label{qva}
   \langle J_V\rangle
   = \langle H(p')|\bar c \!\not\!q \,b|\bar{B}_d(p)\rangle , \qquad
   \langle J_A\rangle
   = \langle H(p')|\bar c\!\not\!q\gamma_5 b\,|\bar{B}_d(p)\rangle .
\end{equation}
In obtaining (\ref{delo8}) we have used the equations of motion for the 
quarks to reduce the operator basis to $J_V$ and $J_A$. 
The form of (\ref{delo8}) is identical 
for pions and longitudinally polarized $\rho$ mesons. The production of  
transversely polarized $\rho$ mesons is power suppressed in 
$\Lambda_{\rm QCD}/m_b$.

In the case of a distribution amplitude $\Phi_L(u)$ that is
symmetric under $u\leftrightarrow \bar u$, which is relevant
for $L=\pi, \rho$, the function $F(u,z)$ appearing in (\ref{delo8}) 
can be compactly written as
\begin{equation}\label{ffsym}
   F(u,z) = 3 \ln z^2 - 7 + f(u,z) + f(u,1/z) ,
\end{equation}
with
\begin{equation}\label{fxzsym}
 f(u,z) = - \frac{u(1-z^2)[3(1-u (1-z^2))+z]}{[1-u(1-z^2)]^2}
    \ln[u(1-z^2)] - \frac{z}{1-u(1-z^2)} .
\end{equation}
The contribution of $f(u,z)$ in (\ref{ffsym}) comes from 
the first two diagrams in Fig.~\ref{fig6} with the gluon coupling to the $b$ 
quark, whereas $f(u,1/z)$ arises from the last two diagrams with the 
gluon coupling to the charm quark. The absorptive part of the amplitude, 
which is responsible for the occurrence 
of strong rescattering phases, arises from $f(u,1/z)$ and can be 
obtained by recalling that $z^2$ is $z^2-i\epsilon$ with $\epsilon>0$ 
infinitesimal. We then have
\begin{equation}
   \frac{1}{\pi}\,\mbox{Im}\,F(u,z)
   = - \frac{(1-u)(1-z^2)[3(1-u (1-z^2))+z]}{[1-u(1-z^2)]^2} 
\end{equation}   

As mentioned above, (\ref{delo8}) is applicable to all decays
of the type $\bar B_d\to D^{(*)+}L^-$, where $L$ is a light
hadron such as a pion or a (longitudinally polarized) $\rho$ meson.
Only the operator $J_V$ contributes to $\bar B_d\to D^+ L^-$, and
only $J_A$ contributes to $\bar B_d\to D^{*+} L^-$. (Due to helicity 
conservation the vector current $B\to D^*$ matrix element contributes
only in conjunction with a transversely polarized $\rho$ meson and hence 
is power suppressed in the heavy-quark limit.) Our final result 
can therefore be written as 
\begin{equation}\label{bdpi18}
\langle D^+ L^-|O_{0,8}|\bar B_d\rangle=
\langle D^+|\bar c\gamma^\mu(1-\gamma_5)b|\bar B_d\rangle
\cdot i f_L q_\mu\int^1_0 du\,T_{0,8}(u,z)\,\Phi_L(u) ,
\end{equation}
where $L=\pi$, $\rho$, and the hard-scattering kernels are
\begin{eqnarray}
\label{t1uz}
   T_0(u,z)&=& 1+O(\alpha^2_s) , \\
\label{t8uz}
   T_8(u,z)&=&\frac{\alpha_s}{4\pi}\frac{C_F}{2N_c} \left[
    - 6\ln\frac{\mu^2}{m_b^2} - B + F(u,z) \right]
    + O(\alpha^2_s) .
\end{eqnarray}
When the $D$ meson is replaced by a $D^*$ meson, the result is 
identical except that $F(u,z)$ in (\ref{t8uz}) must be replaced 
by $F(u,-z)$. Since no order $\alpha_s$ corrections exist for $O_0$, 
the matrix element retains its leading-order factorized form
\begin{equation}\label{o1me}
\langle D^+L^-|O_0|\bar B_d\rangle= i f_L q_\mu\,
\langle D^+|\bar c\gamma^\mu(1-\gamma_5)b|\bar B_d\rangle
\end{equation}
to this accuracy. 
From (\ref{fxzsym}) it follows that $T_8(u,z)$ tends to a 
constant as $u$ approaches the endpoints ($u\to 0$, $1$). 
Therefore the contribution
to (\ref{bdpi18}) from the endpoint region is suppressed,
both by phase space and by the endpoint suppression
intrinsic to $\Phi_L(u)$, which can be represented as
\begin{equation}\label{gpol}
   \Phi_L(u) = 6u(1-u) \left[ 1 + \sum_{n=1}^\infty 
   \alpha_n^L(\mu)\,C_n^{3/2}(2u-1) \right] ,
\end{equation}
vanishing $\sim u$ ($\sim (1-u)$) at the endpoints.
Here $C_1^{3/2}(x)=3x$, $C_2^{3/2}(x)=\frac32(5x^2-1)$, etc. 
are Gegenbauer polynomials. The 
Gegenbauer moments $\alpha_n^L(\mu)$ are nonpertrubative
quantities. They are multiplicatively renormalized and
approach zero as $\mu\to\infty$. 
(The scale dependence of these quantities enters the 
results for the coefficients only at order $\alpha_s^2$, which is 
beyond the accuracy of a NLO calculation.)

As a consequence of the endpoint suppression
the emitted light meson is indeed
dominated by energetic constituents, as required for the self-consistency
of the factorization formula (\ref{bdpi18}).  

Combining (\ref{bheff}), (\ref{bdpi18}), (\ref{t1uz}) and (\ref{t8uz}), we 
obtain our final result for the class-I, non-leptonic $\bar B_d\to D^{(*)+} 
L^-$ decay amplitudes in the heavy-quark limit, and at next-to-leading order 
in $\alpha_s$. The results can be compactly expressed in terms of the matrix
elements of a ``transition operator''
\begin{equation}\label{heffa1}
{\cal T}=\frac{G_F}{\sqrt{2}} V^*_{ud}V_{cb}
\left[ a_1(D L)\, Q_V - a_1(D^* L)\, Q_A\right] ,
\end{equation}
where
\begin{equation}\label{qva2}
Q_V=\bar c\gamma^\mu b\,\otimes\, \bar d\gamma_\mu(1-\gamma_5)u ,
\qquad
Q_A=\bar c\gamma^\mu\gamma_5 b\,\otimes\, \bar d\gamma_\mu(1-\gamma_5)u ,
\end{equation}
and hadronic matrix elements of $Q_{V,A}$ are understood to be evaluated
in factorized form, i.e.\ 
\begin{equation}
   \langle D L|j_1\otimes j_2|\bar B\rangle
   \equiv \langle D|j_1|\bar B\rangle\,\langle L|j_2|0\rangle .
\end{equation} 
Eq.~(\ref{heffa1}) defines the quantities $a_1(D^{(*)} L)$, which include 
the leading ``non-factorizable'' corrections, in a renormalization-scale and 
-scheme independent way. To leading power in $\Lambda_{\rm QCD}/m_b$ these 
quantities should not be interpreted as phenomenological parameters (as is 
usually done), because they are dominated by hard gluon exchange and thus 
calculable in QCD. At next-to-leading order we get
\begin{eqnarray}\label{a1dpi}
   a_1(D L) &=& \frac{N_c+1}{2N_c}\bar C_+(\mu)
    + \frac{N_c-1}{2N_c}\bar C_-(\mu) \nonumber\\
   &&\mbox{}+ \frac{\alpha_s}{4\pi}\frac{C_F}{2N_c}\,C_8(\mu) \left[
    - 6\ln\frac{\mu^2}{m_b^2} + \int^1_0 du\,F(u,z)\,\Phi_L(u)
    \right] , \\
\label{a1dspi}
   a_1(D^* L) &=& \frac{N_c+1}{2N_c}\bar C_+(\mu)
    + \frac{N_c-1}{2N_c}\bar C_-(\mu) \nonumber\\
   &&\mbox{}+ \frac{\alpha_s}{4\pi}\frac{C_F}{2N_c}\,C_8(\mu) \left[
    - 6\ln\frac{\mu^2}{m_b^2} + \int^1_0 du\,F(u,-z)\,\Phi_L(u) \right] .
\end{eqnarray}
These expressions generalize the well-known leading-order formula
\begin{equation}\label{a1lo}
a^{\rm LO}_1=\frac{N_c+1}{2N_c}C^{\rm LO}_+(\mu)
+\frac{N_c-1}{2N_c}C^{\rm LO}_-(\mu) .
\end{equation}
We observe that the scheme dependence, parametrized by $B$, is cancelled
between the coefficient of $O_0$ in (\ref{bheff}) and the matrix
element of $O_8$ in (\ref{bdpi18}). Likewise, the $\mu$ dependence of the 
terms in brackets
in (\ref{a1dpi}) and (\ref{a1dspi}) cancels against the scale dependence of 
the coefficients $\bar C_\pm(\mu)$, ensuring a consistent physical result 
at next-to-leading order in QCD. 

The coefficients $a_1(D L)$ and $a_1(D^* L)$ are seen to be non-universal, 
i.e.\ they are explicitly dependent on the nature of the final-state mesons. 
This dependence enters via the light-cone distribution amplitude $\Phi_L(u)$ 
of the light emission meson and via the analytic form of the hard-scattering 
kernel ($F(u,z)$ vs.\ $F(u,-z)$). However, the non-universality enters only 
at next-to-leading order.

\subsection*{Exercise}

{\it Verify that the coefficients $a_1(D^{(*)}L)$ in
(\ref{a1dpi}) and (\ref{a1dspi}) are independent of the
unphysical renormalization scale $\mu$ up to terms
of ${\cal O}(\alpha^2_s)$.}

\subsubsection{Phenomenological applications for $B\to D\pi$}

An important test of QCD factorization can be performed by
comparing the hadronic decays $\bar B_d\to D^{(*)+}L^-$
with the semileptonic processes $\bar B_d\to D^{(*)+}l^-\nu$.
It is useful to introduce the ratios   
\begin{equation}\label{tfrpi}
   R_L^{(*)} = \frac{\Gamma(\bar B_d\to D^{(*)+} L^-)}
    {d\Gamma(\bar B_d\to D^{(*)+} l^-\bar\nu)/dq^2\big|_{q^2=m^2_L}}
   = 6\pi^2 |V_{ud}|^2 f^2_L\,|a_1(D^{(*)} L)|^2\,X^{(*)}_L ,
\end{equation}
where $X_\rho=X_\rho^*=1$ for a vector meson (because the production
of the lepton pair via a $V-A$ current in semi-leptonic decays is
kinematically equivalent to that of a vector meson with momentum $q$), 
whereas $X_\pi$ and $X_\pi^*$ deviate from 1 only by (calculable) terms 
of order $m_\pi^2/m_B^2$, which numerically are below $1\%$.
The main virtue of (\ref{tfrpi}) is that the $B\to D^{(*)}$
form factors cancel in the ratio. 
The theoretical prediction for the QCD coefficients,
based on QCD factorization 
to leading power and at NLO in perturbative QCD, is
$|a_1(D^{(*)}L)|=1.05$. The uncertainty of this leading-power result
is small, about $\pm 0.01$. The prediction
is to be compared with the experimental results, extracted
from (\ref{tfrpi}), which read 
$|a_1(D^*\pi)|=1.08\pm 0.07$,
$|a_1(D^*\rho)|=1.09\pm 0.10$ and
$|a_1(D^* a_1)|=1.08\pm 0.11$. 
These values show fair agreement with the theoretical
number, albeit within experimental uncertainties that are still large.

Another interesting consideration concerns the
comparison of class-I modes with those of class II and III.
For $B\to D^{(*)}\pi$, all three decay modes, which are
related by isospin, have been measured. A nice
discussion of the present experimental status and its interpretation
in the context of QCD factorization has been given 
by Neubert \& Petrov (2001).
Let us briefly discuss here a few important aspects. 
The experimental status is summarized in Table 3.
\begin{table}[h]
\label{tab:bdpi}
\caption[]{Experimental data for $\bar B\to D^{(*)}\pi$
branching ratios (in units of $10^{-3}$) and extracted
parameters $|C-A|/|T+A|$, $\delta_{TC}$
(see Neubert \& Petrov (2001)).}
\begin{center}
\vspace*{0.4cm}
\begin{tabular}{|c|c|c|}
\hline
 & $\bar B\to D\pi$ & $\bar B\to D^*\pi$\\
\hline
$\bar B^0\to D^{(*)+}\pi^-$ & $3.0\pm 0.4$  &   $2.76\pm 0.21$\\
$\bar B^0\to D^{(*)0}\pi^0$ & $0.27\pm 0.05$  & $0.17\pm 0.05$\\
$B^-\to D^{(*)0}\pi^-$      & $5.3\pm 0.5$  &   $4.6\pm 0.4$\\
\hline
\hline
$|C-A|/|T+A|$ & $0.42\pm 0.05$ & $0.35\pm 0.05$\\
$\delta_{TC}$ & $(56\pm 20)^\circ$ & $(51\pm 20)^\circ$\\
\hline
\end{tabular}
\end{center}
\end{table}
Denoting the basic topologies from 
Figs. \ref{fig:bdpi} (a), (b) and (c), by $T$, $C$ and $A$,
respectively (where the notation refers to
``tree'', ``colour-suppressed tree'' and ``annihilation''), we have  
\begin{eqnarray}\label{asw}
   {\cal A}(\bar B_d\to D^+\pi^-)
   &=& T + A \label{asw+-} \\
   \sqrt2\,{\cal A}(\bar B_d\to D^0\pi^0) 
   &=& C - A  \label{asw00} \\
   {\cal A}(B^-\to D^0\pi^-) &=& T + C  \label{asw0-}
\end{eqnarray}
For later use we may further define the spectator-interaction
contribution to $T$, $T_{spec}$ (see bottom row of Fig. \ref{figlett1}).
A similar decomposition holds for the $\bar B\to D^{*}\pi$ 
modes. Isospin symmetry is reflected in the amplitude relation
${\cal A}(\bar B_d\to D^+\pi^-)$ $+$ $\sqrt2\,{\cal A}(\bar B_d\to D^0\pi^0)$
$=$ ${\cal A}(B^-\to D^0\pi^-)$, which is manifest
in the parametrization of (\ref{asw+-}) -- (\ref{asw0-}).
This means that there are only two 
independent amplitudes, which we can take to be $(T+A)$ and $(C-A)$. 
These amplitudes are complex due to
strong phases from final-state interactions. Only the 
relative phase of the two independent amplitudes is an observable 
and we define $\delta_{TC}$ to be
the relative phase of $(T+A)$ and $(C-A)$.
The phase can then be extracted from the data through
the relation
\begin{equation}\label{deltc}
\cos\delta_{TC} =
\frac{\frac{\tau(\bar B^0)}{\tau(B^-)} 
B(B^-\to D^0\pi^-)-B(\bar B^0\to D^+\pi^-)- 2 B(\bar B^0\to D^0\pi^0)}{
2\sqrt{2}\sqrt{B(\bar B^0\to D^+\pi^-)\, B(\bar B^0\to D^0\pi^0)}}  
\end{equation}

In the usual heavy-quark limit, $m_b\sim m_c\gg\Lambda_{QCD}$, 
only $T$ is calculable. $T_{spec}$, $C$ and $A$ are not, but
they are power suppressed.
It is instructive to consider the alternative limit
$m_b\gg m_c\gg\Lambda_{QCD}$, which allows us to distinguish
between $m_b$ and $m_c$ (Beneke et al. 2000).
In this case, due to $m_c\ll m_b$, also the $D$ becomes
a ``light'' meson and in that respect the process is similar
to $B\to\pi\pi$, where both $T_{spec}$ and $C$ are calculable.
However, since $m_c\gg\Lambda_{QCD}\equiv\Lambda$, the $D$-meson wave
function $\Phi_D(u)$ is highly asymmetric and strongly peaks
at $\bar u\equiv (1-u)\sim\Lambda/m_c$, where $u$ is the
$c$-quark momentum fraction. 
These properties can be used to derive the scaling rules
\begin{equation}\label{atcscal}
\frac{A}{T}\sim\frac{\Lambda}{m_b}\qquad\quad
\frac{T_{spec}}{T}\sim\frac{\Lambda}{m_c}\qquad\quad
\frac{C}{T}\sim\frac{\Lambda}{m_c}
\end{equation}
The amplitude $A$ is still not calculable in this scheme, while
$T$, $T_{spec}$ and $C$ are.
Note that from (\ref{atcscal}) we can recover the two
standard scenarios we have been discussing:
In the heavy-quark limit $m_c\sim m_b$ (\ref{atcscal})
reduces to a simple power suppression $\sim\Lambda/m_b$
for $A$, $T_{spec}$ and $C$ compared to $T$.
On the other hand, if $m_c$ becomes a truly light quark,
corresponding to the case of $B\to\pi\pi$, we count $m_c\sim\Lambda$
and see that both $T_{spec}$ and $C$ are of the same order as $T$,
while $A$ is still power suppressed.

The general scenario $m_b\gg m_c\gg\Lambda_{QCD}$ allows us to
interpret the experimental results in Table 3. We can even
make numerical estimates for $T$, $T_{spec}$ and $C$ based
on the factorization formula for light-light final states (\ref{fform}).
These are somewhat model dependent because $\Phi_D$ is not known
at present. It is not too difficult
to accomodate substantial values $|C-A|/|T+A|\sim 0.2$ -- $0.3$
and $\delta_{TC}\sim 40^\circ$, in qualitative agreement
with Table 3.
Given the special role of the charm quark (not light, but also
not too heavy), the current experimental situation is not
in contradiction with QCD factorization in the large-$m_b$
limit. For a comparison with experiment it is useful
to keep in mind that, according to (\ref{atcscal}), the
suppression of $C$ over $T$ is only $\sim\Lambda/m_c$
(not $\Lambda/m_b$) and that $\delta_{TC}$ can also be substantial.

\subsection*{Exercise}

{\it Derive the relation (\ref{deltc}).}

\subsection{CP violation in $B\to\pi^+\pi^-$ decay}

Hadronic $B$ decays into a pair of light mesons, such
as $B\to \pi K$ or $B\to\pi\pi$, have a very rich
phenomenology. Their main interest lies in their sensitivity
to short-distance flavour physics, including CKM parameters,
CP violation and the search for new physics.
By way of an outlook we mention here the important example
of CP violation in $B\to\pi^+\pi^-$ decay.
The starting point for computing the required decay
amplitudes is the effective Hamiltonian in (\ref{hdb1}).
The needed matrix elements of the operators $Q_i$ can be analyzed
within QCD factorization using (\ref{fform}).
We will not go into the technical details of such an analysis
and the discussion of limitations of the approach, in particular
from power corrections in $\Lambda_{QCD}/m_b$. These can be found
in (Beneke et al. 2001).
Here we just want to present the phenomenological motivation
and to illustrate that a theoretical approach towards a direct
dynamical calculation of hadronic matrix elements will be
very valuable, even if it necessitates some approximations.  

The observable of interest is
the time-dependent CP asymmetry in the decays
$B^0,\bar B^0\to\pi^+\pi^-$, which is sensitive to the $B_d$--$\bar B_d$
mixing phase $e^{-2i\beta}$. We define
\begin{eqnarray}
   A_{\rm CP}^{\pi\pi}(t)
   &=& \frac{B(B^0(t)\to\pi^+\pi^-)
             - B(\bar B^0(t)\to\pi^+\pi^-)}
               {B(B^0(t)\to\pi^+\pi^-)
             + B(\bar B^0(t)\to\pi^+\pi^-)} \nonumber\\
   &=& - S_{\pi\pi} \sin(\Delta m_B\,t)
    + C_{\pi\pi} \cos(\Delta m_B\,t) 
\end{eqnarray}
where
\begin{equation}\label{Spipi}
   S_{\pi\pi} = \frac{2\,\mbox{Im}\,\lambda_{\pi\pi}}
                       {1+|\lambda_{\pi\pi}|^2} \,, \quad
   C_{\pi\pi} = \frac{1-|\lambda_{\pi\pi}|^2}
                       {1+|\lambda_{\pi\pi}|^2} \,, \quad
   \lambda_{\pi\pi} = e^{-2i\beta}\,
    \frac{e^{-i\gamma} + P_{\pi\pi}/T_{\pi\pi}}
         {e^{i\gamma} + P_{\pi\pi}/T_{\pi\pi}} 
\end{equation}
The amplitudes $T_{\pi\pi}$ (``tree'') and 
$P_{\pi\pi}$ (``penguin'') are the
components of the $B\to\pi^+\pi^-$ amplitude corresponding
to the terms in (\ref{hdb1}) involving $\lambda_u$ and $\lambda_c$,
respectively. In the standard phase conventions $\lambda_c$ is
real and $\lambda_u$ has a weak phase $-\gamma$, which has been
factored out above.  
The coefficient $C_{\pi\pi}$, which is a function of 
$\gamma$, represents direct CP violation and is expected to be
small. We shall not discuss it further here.
The mixing-induced asymmetry $S_{\pi\pi}$ depends on $\gamma$ and 
$\beta$. In fact, in the limit where $P_{\pi\pi}/T_{\pi\pi}$ is set to 
zero it follows that $\lambda_{\pi\pi}=e^{-2i(\beta+\gamma)}
=e^{2i\alpha}$, and hence $S_{\pi\pi}=\sin2\alpha$.
In this limit $\lambda_{\pi\pi}$ is just the relative weak
phase between the direct amplitude $B\to\pi^+\pi^-$ and the one
with mixing $B\to\bar B\to\pi^+\pi^-$. All dependence on hadronic
input has canceled in this situation.
In practice, however, $P_{\pi\pi}/T_{\pi\pi}$ is not fully negligible.
It is here that information on hadronic dynamics becomes crucial.
QCD factorization predicts that $P_{\pi\pi}/T_{\pi\pi}$ is
suppressed (either by $\alpha_s$ or
by powers of $\Lambda_{QCD}/m_b$), because $T_{\pi\pi}$ can arise 
at tree level, $P_{\pi\pi}$ only through loops. 
Estimates within this framework give values of about
$0.25$ -- $0.3$. 
\begin{figure}
[htbp]
\centering
\hspace*{0cm}
\includegraphics[width=10cm,height=10cm]{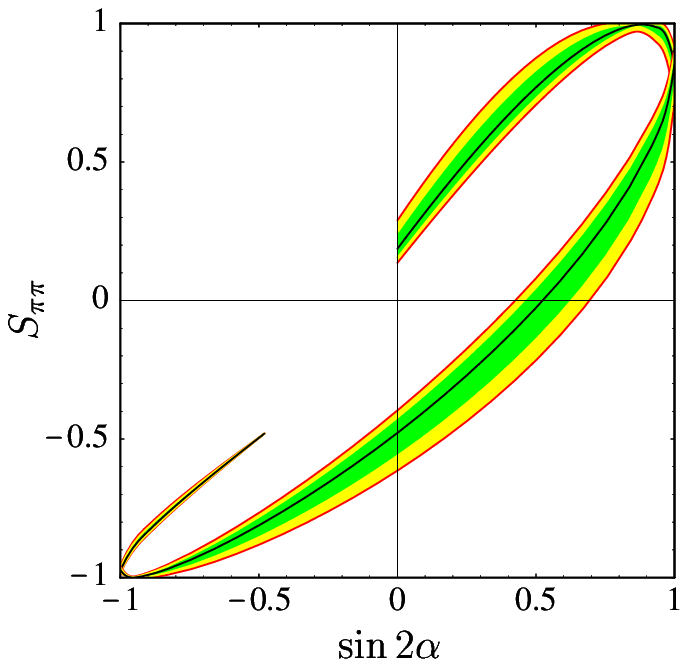}
\vspace*{0cm}
\caption{Relation between $\sin2\alpha$ and the mixing-induced CP asymmetry
$S_{\pi\pi}$, assuming $\sin2\beta=0.48$. The dark band reflects 
parameter variations of the first kind, the light band shows the total 
theoretical uncertainty. The lower portion of the band refers to values 
$45^\circ<\alpha<135^\circ$, the upper one to $0<\alpha<45^\circ$ 
(right branch) or $135^\circ<\alpha<180^\circ-\beta$ (left branch).}
\label{fig:sin2a}
\end{figure}

To illustrate the effect of penguins, we first assume that 
$|V_{ub}/V_{cb}|$ and the weak phase $\beta$ have been determined 
accurately. Then using $\gamma=180^\circ-\alpha-\beta$ the expression 
for $\lambda_{\pi\pi}$ in (\ref{Spipi}) becomes a function of $\alpha$ 
and our prediction for the penguin-to-tree ratio $P_{\pi\pi}/T_{\pi\pi}$.
If we further assume that the unitarity triangle lies in the upper half 
of the $(\bar\rho,\bar\eta)$ plane, then a measurement of $S_{\pi\pi}$ 
determines $\sin2\alpha$ with at most a two-fold discrete ambiguity. 
Figure~\ref{fig:sin2a} shows the relation between the two quantities for 
the particular case where $|V_{ub}/V_{cb}|=0.085$ and 
$\beta=14.3^\circ$, corresponding to $\sin2\beta=0.48$. The dark 
band shows the theoretical uncertainty due to input parameter variations, 
whereas the light band indicates 
the total theoretical uncertainty including estimates of
the effect of power corrections. We observe that
for negative values $\sin2\alpha$ as preferred by the global analysis of 
the unitarity triangle, a measurement of the 
coefficient $S_{\pi\pi}$ could be used to determine $\sin2\alpha$ with 
a theoretical uncertainty of about $\pm 0.1$. Interestingly, for such 
values of $\sin2\alpha$ the ``penguin pollution'' effect enhances the 
value of the mixing-induced CP asymmetry, yielding values of 
$S_{\pi\pi}$ between $-0.5$ and $-1$. Such a large asymmetry should be 
relatively easy to observe experimentally.
\begin{figure}
[ht]
\centering
\hspace*{3cm}
\includegraphics[width=20cm,height=10cm]{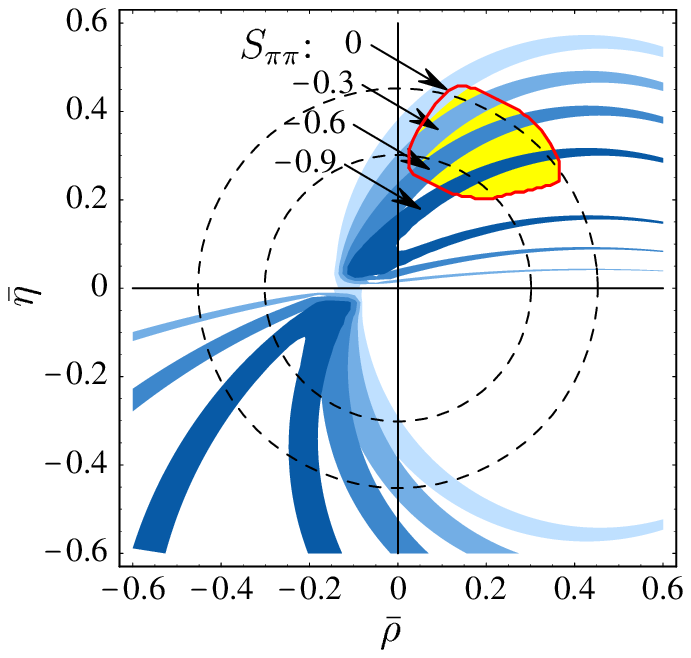}
\vspace*{-0.4cm}
\caption{Allowed regions in the $(\bar\rho,\bar\eta)$ plane corresponding to
constant values of the mixing-induced asymmetry $S_{\pi\pi}$ assuming the 
Standard Model. 
The widths of the bands reflect the total theoretical uncertainty. The 
corresponding bands for positive values of $S_{\pi\pi}$ are obtained by 
a reflection about the $\bar\rho$ axis. The light circled area in the
left-hand plot shows the allowed region obtained from the standard 
global fit of the unitarity triangle (H\"ocker et al. 2001).}
\label{fig:fancy}
\end{figure}

Although it illustrates nicely the effect of ``penguin pollution'' on
the determination of $\sin2\alpha$, Figure~\ref{fig:sin2a} is not the
most appropriate way to display the constraint on the unitarity triangle 
implied by a measurement of $S_{\pi\pi}$. In general, there is a 
four-fold discrete ambiguity in the determination of $\sin2\alpha$, 
which we have reduced to a two-fold ambiguity by assuming that the 
triangle lies in the upper half-plane. Next, and more importantly, we 
have assumed that $|V_{ub}/V_{cb}|$ and $\beta$ are known with precision,
whereas $\alpha$ is undetermined. However, in the Standard Model 
$|V_{ub}/V_{cb}|$ and the angles $\alpha,\beta,\gamma$ are all functions 
of the Wolfenstein parameters $\bar\rho$ and $\bar\eta$. It is thus more 
appropriate to represent the constraint implied by a measurement of 
$S_{\pi\pi}$ as a band in the $(\bar\rho,\bar\eta)$ plane. To this end, 
we write
\begin{equation}
   e^{\mp i\gamma}
   = \frac{\bar\rho\mp i\bar\eta}{\sqrt{\bar\rho^2+\bar\eta^2}}
     \quad
   e^{-2i\beta}
   = \frac{(1-\bar\rho)^2-\bar\eta^2-2i\bar\eta(1-\bar\rho)}
          {(1-\bar\rho)^2+\bar\eta^2}  \quad
   \frac{P_{\pi\pi}}{T_{\pi\pi}}
   = \frac{r_{\pi\pi}\,e^{i\phi_{\pi\pi}}}
          {\sqrt{\bar\rho^2+\bar\eta^2}} 
\end{equation}
where $r_{\pi\pi}\,e^{i\phi_{\pi\pi}}$ is independent 
of $\bar\rho$ and $\bar\eta$. We now insert these relations into 
(\ref{Spipi}) and draw contours of constant $S_{\pi\pi}$ in the
$(\bar\rho,\bar\eta)$ plane. The result is shown by the bands in 
Figure~\ref{fig:fancy}. 
The widths of the bands 
reflect the total theoretical uncertainty (including power corrections).
For clarity we show only bands for negative values of $S_{\pi\pi}$; those
corresponding to positive $S_{\pi\pi}$ values can be obtained by a 
reflection about the $\bar\rho$ axis (i.e., $\bar\eta\to-\bar\eta$). Note
that even a rough measurement of $S_{\pi\pi}$ would translate into a
rather narrow band in the $(\bar\rho,\bar\eta)$ plane, 
which intersects the ring representing the 
$|V_{ub}/V_{cb}|$ constraint at almost right angle. 
In a similar way, the constraint
is also quite robust against hadronic uncertainties.
Even the approximate knowledge of hadronic
matrix elements, as provided by QCD factorization, 
will therefore be very valuable and can  
lead to powerful constraints on the Wolfenstein parameters.

\section{Summary}

In these lectures we have discussed the theory of heavy
quarks, focussing on the important case of $B$ physics.
(All methods relying on the heavy-quark limit
could in principle be applied to charmed hadrons
as well, but they are in most cases much less reliable
due to the smaller value of the charm mass.) 
We shall conclude by summarizing the key points.

\begin{itemize}
\item 
A crucial and general idea for dealing with QCD effects
is the {\it factorization of short-distance and 
long-distance dynamics}. We have encountered this
principle in many different forms and applications: 
\begin{itemize}
\item
The OPE to construct the effective weak Hamiltonians
(${\cal H}^{\Delta B=1,2}_{eff}$) factorizes the
short-distance scales of order $M_W$, $m_t$ from
the scales of order $m_b$.
\item
The heavy-quark scale
$m$ treated as a short-distance scale can be factorized further
from the intrinsic long-distance scale of QCD, $\Lambda_{QCD}$.
This leads to a systematic expansion of observables simultaneously
in $1/m$ and $\alpha_s(m)$ with often very important 
simplifications. The precise formulation of this class of factorization
depends on the physical situation and can take the form
of HQET, LEET, HQE or QCD factorization in exclusive hadronic $B$ decays.
\end{itemize}
\item
HQET exhibits the spin-flavour symmetry of QCD in the heavy-quark limit,
which allows us to relate different form factors, and makes the
$m_Q$ dependence explicit. Examples of typical applications
are $B\to D^{(*)}l\nu$ or $f_B$.
\item
HQE is a theory for inclusive $B$ decays. It justifies
the ``parton model'' and allows us to study the nonperturbative
power corrections. This is of great use for
processes as $B\to X_{c,u}l\nu$, $B\to X_s\gamma$,
$B\to X_s l^+l^-$, and the lifetimes
of $b$-flavoured hadrons.
\item
QCD factorization, finally, refers to a framework for
analyzing exclusive hadronic $B$ decays with a fast light meson
as for instance $B\to D\pi$, $B\to\pi\pi$, $B\to\pi K$ and
$B\to V\gamma$.
\end{itemize}

With these tools at hand we are in a good position to make
full use of the rich experimental results in the physics of heavy 
flavours. We can determine fundamental parameters of the
flavour sector, such as $V_{ub}$, $V_{cb}$, $V_{td}/V_{ts}$,
$\eta$ and $\sin 2\alpha$, and probe electroweak dynamics
at the quantum level through $b\to s\gamma$ or $B$ -- $\bar B$ mixing.
This will enable us to thoroughly test the standard model
and to learn about new structures and phenomena that are
yet to be discovered.

\section*{Acknowledgments}

I thank the organizers of SUSSP 55 for inviting me to 
this most enjoyable and perfectly orchestrated
Summer School and all the participants for contributing to
the extraordinary atmosphere at St. Andrews. 
Section 6 of these lectures is based on work done together
with Martin Beneke, Matthias Neubert and Chris Sachrajda,
to whom I am grateful for a very pleasant and fruitful collaboration.

\section*{References}
\frenchspacing
\begin{small}

\reference{Albrecht~H et al. [ARGUS Collaboration], 1987,
Observation of $B^0$ - $\bar B^0$ mixing,
\textit{Phys. Lett. B} \textbf{192} 245.}

\reference{Altarelli~G, Curci~G, Martinelli~G and Petrarca~S, 1981,
QCD nonleading corrections to weak decays as an application of 
regularization by dimensional reduction,
\textit{Nucl. Phys. B} \textbf{187} 461.}

\reference{Anikeev~K et al., 2002,
\textit{$B$ Physics at the Tevatron: Run II and Beyond},
FERMILAB-Pub-01/197, [\textit{hep-ph}/0201071].}

\reference{Bauer~C~W, Fleming~S, Pirjol~D and Stewart~I~W, 2001a,
An effective field theory for collinear and soft gluons: 
heavy to light decays,
\textit{Phys. Rev. D} \textbf{63} 114020
[\textit{hep-ph}/0011336].}

\reference{Bauer~C~W, Ligeti~Z and Luke~M~E, 2001b,
Precision determination of $|V_{ub}|$ from inclusive decays,
\textit{Phys. Rev. D} \textbf{64} 113004
[\textit{hep-ph}/0107074].}

\reference{Bauer~C~W, Pirjol~D and Stewart~I~W, 2001c,
A proof of factorization for $B\to D\pi$,
\textit{Phys. Rev. Lett.}  \textbf{87} 201806
[\textit{hep-ph}/0107002].}

\reference{Beneke~M, Buchalla~G, Neubert~M and Sachrajda~C~T, 1999,
QCD factorization for $B\to\pi\pi$ decays: 
Strong phases and CP  violation in the heavy quark limit,
\textit{Phys. Rev. Lett.}  \textbf{83} 1914
[\textit{hep-ph}/9905312].}

\reference{Beneke~M, Buchalla~G, Neubert~M and Sachrajda~C~T, 2000,
QCD factorization for exclusive, non-leptonic B meson decays: 
General  arguments and the case of heavy-light final states,
\textit{Nucl. Phys. B} \textbf{591} 313
[\textit{hep-ph}/0006124].}

\reference{Beneke~M, Buchalla~G, Neubert~M and Sachrajda~C~T, 2001,
QCD factorization in $B\to\pi K$, $\pi\pi$ decays and 
extraction of  Wolfenstein parameters,
\textit{Nucl. Phys. B} \textbf{606} 245
[\textit{hep-ph}/0104110].}

\reference{Beneke~M and Feldmann~T, 2001,
Symmetry-breaking corrections to heavy-to-light B meson form factors 
at large recoil,
\textit{Nucl. Phys. B} \textbf{592} 3
[\textit{hep-ph}/0008255].}

\reference{Bigi~I~I, Blok~B, Shifman~M~A, Uraltsev~N and Vainshtein~A~I, 1994,
Nonleptonic decays of beauty hadrons: From phenomenology to theory,
\textit{hep-ph}/9401298.}

\reference{Bigi~I~I, Shifman~M~A and Uraltsev~N, 1997,
Aspects of heavy quark theory,
\textit{Ann. Rev. Nucl. Part. Sci.} \textbf{47} 591
[\textit{hep-ph}/9703290].}

\reference{Bigi~I~I and Uraltsev~N, 2001,
A vademecum on quark hadron duality,
\textit{Int. J. Mod. Phys. A} \textbf{16} 5201
[\textit{hep-ph}/0106346].}

\reference{Bigi~I~I, Uraltsev~N~G and Vainshtein~A~I, 1992,
Nonperturbative corrections to inclusive beauty and charm decays: 
QCD versus phenomenological models,
\textit{Phys. Lett. B} \textbf{293} 430;
\textit{Erratum-ibid. B} \textbf{297} (1993) 477
[\textit{hep-ph}/9207214].}

\reference{Bjorken~J~D, 1989,
Topics in B physics,
\textit{Nucl. Phys. Proc. Suppl.}  \textbf{11} 325.}

\reference{Blok~B, Shifman~M~A and Zhang~D~X, 1998,
An illustrative example of how quark-hadron duality might work,
\textit{Phys. Rev. D} \textbf{57} 2691;
\textit{Erratum-ibid. D} \textbf{59} 019901
[\textit{hep-ph}/9709333].}

\reference{Buchalla~G, Buras~A~J and Lautenbacher~M~E, 1996,
Weak decays beyond leading logarithms,
\textit{Rev. Mod. Phys.} \textbf{68} 1125
[\textit{hep-ph}/9512380].}

\reference{Buras~A~J, 1998,
Weak Hamiltonian, CP violation and rare decays,
[\textit{hep-ph}/9806471].}

\reference{Buras~A~J and Weisz~P~H, 1990,
QCD nonleading corrections to weak decays in dimensional 
regularization and 't Hooft-Veltman schemes,
\textit{Nucl. Phys. B} \textbf{333} 66.}

\reference{Charles~J, Le Yaouanc~A, Oliver~L, P\`{e}ne~O and Raynal~J-C, 1999,
Heavy-to-light form factors in the final hadron large energy limit of QCD,
\textit{Phys. Rev. D} \textbf{60} 014001
[\textit{hep-ph}/9812358].}

\reference{Chay~J, Georgi~H and Grinstein~B, 1990,
Lepton energy distributions in heavy meson decays from QCD,
\textit{Phys. Lett. B} \textbf{247} 399.}

\reference{Dugan~M~J and Grinstein~B, 1991,
QCD basis for factorization in decays of heavy mesons,
\textit{Phys. Lett. B} \textbf{255} 583.}

\reference{Gaillard~M~K and Lee~B~W, 1974,
Rare decay modes of the $K$ - mesons in gauge theories,
\textit{Phys. Rev. D} \textbf{10} 897.}

\reference{Harrison~P~F and Quinn~H~R [BABAR Collaboration], 1998,
\textit{The BaBar Physics Book: Physics at an Asymmetric B Factory},
SLAC-R-0504.}

\reference{H\"ocker~A, Lacker~H, Laplace~S and Le Diberder~F, 2001,
A new approach to a global fit of the CKM matrix,
\textit{Eur. Phys. J. C} \textbf{21} 225
[\textit{hep-ph}/0104062]; http://ckmfitter.in2p3.fr}

\reference{Isgur~N and Wise~M~B, 1989,
Weak decays of heavy mesons in the static quark approximation,
\textit{Phys. Lett. B} \textbf{232} 113;
1990, Weak transition form-factors between heavy mesons,
\textit{Phys. Lett. B} \textbf{237} 527.}

\reference{Ligeti~Z, 2001,
New results on flavor physics,
\textit{hep-ph}/0112089.}

\reference{Manohar~A~V and Wise~M~B, 2000,
\textit{Heavy Quark Physics},
Cambridge Monogr. Part. Phys. Nucl. Phys. Cosmol., Vol. \textbf{10}.}

\reference{Neubert~M, 1994,
Heavy quark symmetry,
\textit{Phys. Rept.} \textbf{245} 259
[\textit{hep-ph}/9306320].}

\reference{Neubert~M and Petrov~A~A, 2001,
Comments on color suppressed hadronic B decays,
\textit{Phys. Lett. B} \textbf{519} 50
[\textit{hep-ph}/0108103].}

\reference{Shifman~M~A, 2000,
Quark-hadron duality,
\textit{hep-ph}/0009131.}

\reference{Sterman~G and Stoler~P, 1997,
Hadronic form factors and perturbative QCD,
\textit{hep-ph}/9708370.}

\end{small}


\end{document}